\documentclass[aps, prx, reprint, superscriptaddress, notitlepage, floatfix, nofootinbib]{revtex4-2}

\usepackage[utf8]{inputenc}
\usepackage{amsmath,amssymb,amsthm,amscd,latexsym,epsfig,bm}

\usepackage{physics}
\usepackage{dsfont}
\usepackage{graphicx}
\usepackage{xcolor}
\usepackage{graphicx}

\usepackage{tensor}

\usepackage{cancel}
\usepackage{color}
\usepackage{tikz}
\usepackage{mathrsfs}
\usepackage{relsize}
\usepackage[linktocpage]{hyperref}
\usepackage[all]{xy}
\hypersetup{colorlinks=true,citecolor=blue,linkcolor=blue, urlcolor=blue, breaklinks=true}
\usepackage{verbatim}
\usepackage{braket}
\usepackage{tikz-cd}

\usepackage{bm} 
\usepackage{subfigure} 
\usepackage{environ}
\usepackage{url}
\usepackage[margin=1in]{geometry}
\usepackage[export]{adjustbox}
\usepackage{float}

\usepackage{framed}
\usepackage{booktabs}
\usepackage{longtable}
\usepackage{amstext}
\usepackage{array}
\newcolumntype{L}{>{$}l<{$}} 
\newcolumntype{C}{>{$}c<{$}} 

\def\beq{\begin{equation}}
\def\eeq{\end{equation}}
\def\be{\begin{equation}}
\def\ee{\end{equation}}

\def\ket#1{\vert #1 \rangle}
\def\bra#1{\langle #1 \vert}

\def\ip#1#2{\langle #1 \vert #2 \rangle}
\def\me#1#2#3{\langle #1 \vert #2 \vert #3 \rangle}
\newcommand{\zz}{\mathbb{Z}_2}
\newcommand{\z}{\mathbb{Z}}

\newcommand{\spd}[1]{(#1+1)$d$}

\def\f2{{\mathbb F}_2}

\def\U{\mathrm{U}(1)}

\theoremstyle{plain}
\theoremstyle{plain}

\providecommand{\theoremname}{Theorem}
\providecommand{\theoremtextname}{Theorem}

\theoremstyle{plain}
\providecommand{\propositionname}{Proposition}

\begin{document}

\title{Fermionic quantum criticality through the lens of topological holography}

\author{Sheng-Jie Huang}
\affiliation{Max Planck Institute for the Physics of Complex Systems, N\"othnitzer Str. 38, 01187 Dresden, Germany}

\date{\today}

\begin{abstract} 

    We utilize the topological holographic framework to characterize and gain insights into the nature of quantum critical points and gapless phases in fermionic quantum systems. Topological holography is a general framework that describes the generalized global symmetry and the symmetry charges of a local quantum system in terms of a slab of a topological order, termed as the symmetry topological field theory (SymTFT), in one higher dimension. In this work, we consider a generalization of the topological holographic picture for \spd{1} fermionic quantum phases of matter. We discuss how spin structures are encoded in the SymTFT and establish the connection between the formal fermionization formula in quantum field theory and the choice of fermionic gapped boundary conditions of the SymTFT. We demonstrate the identification and the characterization of the fermionic gapped phases and phase transitions through detailed analysis of various examples, including the fermionic systems with $\zz^{F}$, $\zz \times \zz^{F}$, $\z_{4}^{F}$, and the fermionic version of the non-invertible $\text{Rep}(S_{3})$ symmetry. Our work uncovers many exotic fermionic quantum critical points and gapless phases, including two kinds of fermionic symmetry enriched quantum critical points, a fermionic gapless symmetry protected topological (SPT) phase, and a fermionic gapless spontaneous symmetry breaking (SSB) phase that breaks the fermionic non-invertible symmetry.

\end{abstract}

\maketitle

\tableofcontents

\section{Introduction}

Understanding the phases and transitions in condensed matter physics stands as a central theme in condensed matter physics. Landau paradigm provides a unified frameworks that characterized phases and phase transitions by the symmetry and the patterns of their breaking. However, recent developments in the field have led to the discovery of novel phases of matter that cannot be described by Landau theory such as topological orders with long-rang entanglement \cite{Wen1990TO} as well as phase transitions that are forbidden by the Landau theory \cite{Senthil2004DQCP}. 

In developing theoretical frameworks that characterizes and classifies phase and phase transitions, symmetry has been played a very important role as an organizing principle. It provides valuable insights into classifications and also yields significant physical implications, including the emergence of conservation laws and constraints on low-energy dynamics. The generalization of the notion of symmetry builds on a fact that there is an one-to-one correspondence between the global symmetry described by a symmetry group $G$ and codimension-$1$ topological defects satisfying group multiplication fusion rules. There are however various different kinds of topological defects that could have higher-codimensions, satisfying non-invertible fusion rules, and have a non-trivial interplay among themselves, which have been conjectured to have a higher categorical description \cite{Etingof2009,Kapustin2010,Kong2014,Kong2015,Douglas2018,Kong2020nd,Johnson-Freyd2022,Bhardwaj2022,Bartsch2023,Kong2024}. The modern perspective of the symmetry is that every topological defect should be viewed as a generalized global symmetry \cite{Gaiotto2015,Cordova2022,McGreevy2023}. 

Recently, a novel topological holographic framework has emerged, offering a holographic perspective on symmetry. At its core, this framework encodes symmetry data within a topological order residing in one higher dimension, thereby extending earlier developments within the context of $(1+1)d$ rational CFTs \cite{Fuchs2002TFT1,Fuchs2004TFT2,Fuchs2004TFT3,Fuchs2005TFT4}. The central picture is given by the “sandwich” construction, where we can view the $D$ spacetime dimensional quantum system of interest as a slab of topological order in $D+1$ dimensions with appropriately chosen boundary conditions. This topological order is commonly called  a \emph{symmetry topological field theory} (SymTFT) or a \emph{symmetry topological order} (SymTO)~\cite{Kong2015,Kong2017,Freed2018,Kong2018,Thorngren2019,Bhardwaj2020,Kong2020,Kong2020bdy1,Ji2020,Gaiotto2021,Lichtman2021,Apruzzi2021,Ji2021,Kong2021bdy2,Freed2022,Kaidi2022,Moradi2022,Chatterjee2022local,Chatterjee2022,Chatterjee2022max,Kong2022ql,Kong20221d,Lin2023,Kaidi2023,Benini2023,vanBeest2023,Bhardwaj2023,Bhardwaj2023gapped-1,Bhardwaj2023gapped-2,Huang2023symtft,Bhardwaj2024gspt}. This approach essentially separates the generalized symmetry and the generalized symmetry charges from the dynamics of a quantum system and encode such information in the bulk topological order. It can describe both gapped and gapless phases in a unified framework, and also allows one to make non-perturbative statements about phases, phases transitions, and dualities. Recent developments towards physical applications include the generalization to continuous or non-finite symmetries~\cite{Brennan2024,Apruzzi2024,Antinucci2024,Bonetti2024}, the studying of the gapped phases with non-invertible fusion category symmetries~\cite{Bhardwaj2023gapped-1,Bhardwaj2023gapped-2}, understanding exotic quantum critical points and gapless SPT phases in bosonic systems~\cite{Chatterjee2022,Huang2023symtft,Potter2023,Bhardwaj2023club,Bhardwaj2024gspt}, and constructing lattice models with non-invertible symmetries~\cite{Bhardwaj2024latticelong,Bhardwaj2024latticeshort,Chatterjee2024s3}.

Our main focus in this work is to study quantum criticality in fermionic systems in \spd{1} from the point of view of topological holography. Traditionally, one invaluable tool in tackling this challenge is the bosonization and the Jordan–Wigner transformation. This technique provide a bridge between the fermions and bosons, facilitating the exploration of fermionic systems by mapping them onto more tractable bosonic counterparts. However, identifying the correct conformal field theory (CFT) description of a phase transition remains to be a non-trivial task, which often required numerical simulations. It's thus desirable to develop a theoretical framework that can help to identify possible CFTs of a phase transition. Previous works have applied the SymTFT to characterize phase transitions and gapless SPT phases for bosonic systems~\cite{Chatterjee2022,Huang2023symtft,Potter2023,Bhardwaj2023club,Bhardwaj2024gspt}. However, much less is known regarding the application of this framework to phase transitions and gapless phases for fermionic systems and the resulting physical implications. 

In this work, we apply the idea of topological holography to characterize fermionic gapped phases and identify possible CFT descriptions for the phase transitions. In order to proceed, we need a generalization of the sandwich construction for fermionic systems in the topological holography framework. Recall the central picture of the “sandwich” construction is that we can view the \spd{1} quantum system of interest as a slab of \spd{2} topological order with appropriately chosen boundary conditions. The left boundary for bosonic systems is chosen to be a gapped topological boundary, on which some bosonic anyons in the bulk SymTFT condense, and the confined anyon lines on the left boundary implement the global symmetry of the original \spd{1} system. To apply the idea of topological holography to fermoinic systems, one needs to choose a fermionic gapped boundary condition involving the condensation of the fermionic anyons~\cite{Wan2017,Aasen2019,Lou2021} on the symmetry boundary. This generalized SymTFT picture for fermions was proposed in Ref.~\cite{Gaiotto2016} (see also Ref.~\cite{Freed2022,Debray2023,BoyleSmith2024,Kantaro_pirsa_PIRSA:24030089} for recent developments). The confined anyons on the fermion condensed boundary implement the global symmetry of the fermoinic system. By choosing a physical boundary condition on the right, the sandwich produces the \spd{1} fermionic system that we would like to study. In this work, we show how to apply this picture to study fermionic quantum critical points and gapless SPT phases. In applying the SymTFT framework to study phase transitions, we always need some input critical theories, which lives at the physical boundary, that describe the transitions between the topological gapped boundaries of the SymTFT. Such input critical theories usually can be obtained from some familiar critical points such as the symmetry breaking transitions or more systematically by performing the ``holographic modular bootstrap"~\cite{Ji2019,Ji2021,Chatterjee2023holo}. Many fermionic quantum critical points can be obtained with very little input.  

\subsection{Summary of results}
We summarize the main results of this work below. In Sec.~\ref{sec:fermionization}, we discuss the fermionization in the topological holography framework. In general, performing the fermionization corresponds to change the symmetry boundary to a fermionic gapped boundary conditions such that some fermionic anyons in the bull SymTFT are condensed. We review the fermionic gapped boundaries and super Lagrangian algebra in Appendix~\ref{app:superL}. We establish the connection between the formal fermionization formula that is commonly used in quantum field theory and the choice of the fermionic gapped boundary conditions on the symmetry boundary in Eq.~\eqref{eq:Znsns}-\eqref{eq:Zrr}. We explain how spin structures are encoded in the SymTFT, and give explicit general formulas in Eq.~\eqref{eq:ztwistanyon-f} for fermionic partition functions on the torus with different spin structures in terms of the partition functions in the SymTFT anyon basis. 

We discuss one of the simplest example in Sec.~\ref{sec:z2f}: Fermionic systems with $\zz^{F}$ fermion parity symmetry. We show that the bulk SymTFT is given by the $\zz$ toric code and the symmetry boundary is the fermion $f$ condensed boundary. We explain how to identify the only two possible gapped phases, trivial and the Kitaev phases, with the bosonic gapped boundaries of the $\zz$ toric code by investigating the behavior of the twisted parition functions on the torus. 

We then apply the topological holographic picture to understand the transition between the trivial and the Kitaev phases. The simplest example is given by the free Majorana CFT, which is obtained by assuming that there is an Ising CFT living at the physical boundary of the $\zz$ toric code SymTFT. The partition functions with different spin structures are calculated explicitly in this framework. We can also take the input to be any minimal model CFTs sitting at the physical boundary and the sandwich construction will produce the fermionic minimal model CFTs. The general discussion for the fermoinic minimal models is presented in Appendix~\ref{app:fminimal}.

We discuss fermionic systems with $\zz \times \zz^{F}$ symmetry in Sec.~\ref{sec:z2z2f} with the SymTFT to be the $\zz \times \zz$ toric code. We choose the symmetry boundary to be one of the fermion condensed boundary that corresponds to fermionizing with respecting to the diagonal $\zz$ subgroup in the $\zz \times \zz$ symmetry from the bosonic side. All the gapped phases are identified with the bosonic gapped boundary by examining the twisted partition functions with different spin structures. 

We discuss fermionic quantum critical points with $\zz \times \zz^{F}$ symmetry in Sec.~\ref{sec:z2z2f} and Sec.~\ref{sec:fseqcp}. We consider the input boundary critical theory to be the $\text{Ising}$ and the $\text{Ising}^{2}$ CFTs. The $\text{Ising}$ CFT describes the transition between the gapped boundaries with the Lagrangian algebra $1 \oplus m_{1} \oplus e_{2} \oplus m_{1}e_{2}$ and the $m$-condensed boundary $1 \oplus m_{1} \oplus m_{2} \oplus m_{1}m_{2}$. The $\text{Ising}^{2}$ CFT describes the transition between the $e$- and the $m$-condensed boundaries. With these two input CFTs, we obtain various fermionic quantum critical points between the gapped phases. The fermionic quantum critical points discussed in Sec.~\ref{sec:z2z2f} are described by common fermionic CFTs. We give a detailed discussion on how to identify these CFTs and calculate their partition functions explicitly. 

Some transitions that could happen in the $\zz \times \zz^{F}$ fermionic chains are described by the symmetry enriched quantum critical points, where there are non-trivial SPT signatures in the twisted sectors, which are discussed in Sec.~\ref{sec:fseqcp}. These exotic quantum critical points generally happens at the transition between a SSB phase, where the bosonic symmetry is broken, and a fermionic SPT phase. We find that there are two types of the fermionic symmetry enriched CFTs. The first type of the fermionic symmetry enriched CFTs can be understood as the stacking of the usual SSB transitions with an fermionic SPT state  ($\textbf{T}$ operation). The twisted sectors have addition phase factors corresponding to the SPT invariants. The second type of the fermionic symmetry enriched CFTs are obtained by performing a ``twisted gauging" to the usual ones, where the twisted gauging procedure can be understood as a combination of gauging and stacking a SPT state ($\textbf{T}\textbf{S}$ operation). We find a non-trivial example that describes the transition between the $\zz$ SSB phase and the gapped phase with coexisting Kitaev and the $\zz \times \zz^{F}$ fermionic SPT phases (we refer it as the ``Kitaev + GW" phase later). It belongs to the universality class of the $\text{Ising} \times \text{Majorana}$ CFT. However, there are non-trivial phenomena in the twisted sectors such as the insertion of a $\zz$ defect inducing the change of spin structure. 

In Sec.~\ref{sec:z4f} we discuss fermionic systems with $\z_{4}^{F}$ symmetry. We show that such systems can be obtained by fermionizing with respecting to the $\zz$ subgroup of a $\z_{4}$ symmetry on the bosonic side, or by fermionizing with respecting to one of the $\zz$ subgroup of an anomalous $\zz \times \zz$ symmetry with the mixed anomaly. We identify all the gapped phases and the critical points by using the SymTFT picture. We also discuss a fermionic example of the intrinsically gapless SPT (igSPT) phases with the microscopic $\z_{4}^{F}$ symmetry such that the low-energy sector realizes the bosonic $\zz$ anomaly in $H^{3}(\zz,U(1))$.

In Sec.~\ref{sec:fnoninv} we discuss fermionic systems with non-invertible symmetry. We focus on a fermionic non-invertible symmetry which is obtained by fermoinizing the $\zz$ subgroup in the $\text{Rep}(S_{3})$ symmetry. The resulting symmetry is described by a super fusion category that is generated by the $\zz^{F}$ and a single non-invertible symmetry. It shares the same fusion rule as the $\text{Rep}(S_{3})$ symmetry (refer to as $\text{SRep}(\mathcal{H}_{S_{3}})$ later). We first classify the gapped phases with the $\text{SRep}(\mathcal{H}_{S_{3}})$ symmetry by using the topological holography where the SymTFT is given by the $S_{3}$ gauge theory. There are four possible gapped phases depending on whether the non-invertible symmetry is broken, as well as whether the system is in the Kitaev phase or not. Some of the critical points are governed by non-trivial critical theories. We find the transition between the $\text{SRep}(\mathcal{H}_{S_{3}})$ symmetric trivial and the ``$\text{SRep}(\mathcal{H}_{S_{3}})/\mathbb{Z}_{2}^{F} \ \text{SSB} + \text{Kitaev}$" phases \footnote{We use ``$\text{SRep}(\mathcal{H}_{S_{3}})/\mathbb{Z}_{2}^{F}$ \text{SSB}" to denote the phase that breaks the non-invertible symmetry in $\text{SRep}(\mathcal{H}_{S_{3}})$ spontaneously but preserves the $\zz^{F}$ symmetry} can be described by the fermionic $m=5$ minimal model CFT. The non-trivial automorphism in the $S_{3}$ gauge theory that exchanges the non-abelian charge $C$ and flux $F$ guarantees that the the transition between the $\text{SRep}(\mathcal{H}_{S_{3}})$ symmetric Kitaev and the $\text{SRep}(\mathcal{H}_{S_{3}})/\mathbb{Z}_{2}^{F} \ \text{SSB}$ phases are described by the same fermionic CFT. Another non-trivial critical theory happens at the transition between the $\text{SRep}(\mathcal{H}_{S_{3}})/\mathbb{Z}_{2}^{F} \ \text{SSB}$ and the $\text{SRep}(\mathcal{H}_{S_{3}})/\mathbb{Z}_{2}^{F} \ \text{SSB} + \text{Kitaev}$ phases. This critical theory is the ``sum" of the Ising and the Majorana CFTs. This is an example of the \emph{gapless SSB} phases introduced in Ref.~\cite{Bhardwaj2024hasse}, where the spontaneous breaking of the non-invertible symmetry leads to two superselection sectors that support two distinct CFTs. 

The structure of the paper is as follows. In Sec.~\ref{sec:symtftboson}, we review the central idea of the topological holography for bosonic systems.  We discuss the generalization of such framework to fermionic systems in Sec.~\ref{sec:fermionization} and establish the connection between the formal fermionization formula in quantum field theory and the choice of the fermionic gapped boundary conditions of the SymTFT. Fermionic gapped boundaries are reviewed in Appendix~\ref{app:superL}. We also explain how spin structures are encoded in the SymTFT. In Sec.~\ref{sec:z2f}, we discuss fermionic systems with $\zz^{F}$ fermion parity symmetry as a simple example. We show how to identify the gapped phases and how to use the SymTFT to characterize the phase transition. The simplest critical theory is given by the free Majorana CFT, which is discussed in Sec.~\ref{sec:freemajorana}. Other fermionic minimal models are discussed in Appendix~\ref{app:fminimal}. In Sec.~\ref{sec:z2z2f}, we consider ferminic chains with $\zz \times \zz^{F}$ symmetry and discuss fermionic quantum critical points that are described by commnn fermionic CFTs in Sec.~\ref{sec:z2z2fqcp}. In Sec.~\ref{sec:fseqcp}, we focus on the fermionic symmetry enriched quantum critical points in the $\zz \times \zz^{F}$ fermionic systems. We find that there are two
types of the fermionic symmetry enriched quantum critical points with distinct signatures in the twisted sectors. In Sec.~\ref{sec:z4f}, we discuss fermionic systems with $\z_{4}^{F}$ symmetry and to obtain such symmetry by fermionizing bosonic systems. We then classify the gapped phases, and characterize the phase transitions and a gapless SPT phase. In Sec.~\ref{sec:fnoninv} we discuss fermionic systems with non-invertible symmetry. We focus on a fermionic non-invertible $\text{SRep}(\mathcal{H}_{S_{3}})$ symmetry, which is obtained by fermoinizing the $\zz$ subgroup in the $\text{Rep}(S_{3})$ symmetry. We classify all possible gapped phases and discuss their phase transitions. We conclude with a discussion in Sec.~\ref{sec:discussion}.


\section{Review: topological holography for bosonic systems}
\label{sec:symtftboson}

\begin{figure}
	\centering
	\includegraphics[width=0.45\textwidth]{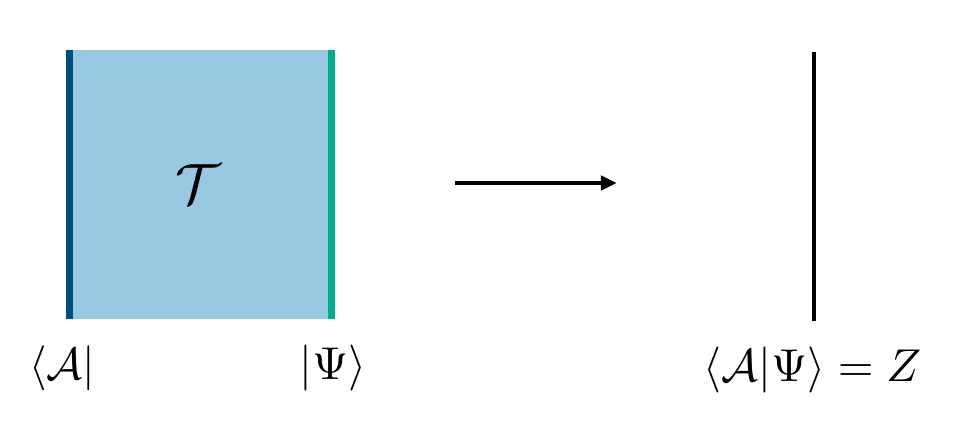}
	\caption{The sandwich picture in the topological holography. A \spd{1} theory can be viewed as a sandwich built from a topological order $\mathcal{T}$ in the bulk with a topological gapped boundary conditions on the left and a potentially non-topological boundary condition on the right.}
	\label{fig:sandwich}
\end{figure}

Here we review the basic idea of the topological holography for bosonic systems. The essential idea of the topological holography is to encode the symmetry data and the corresponding symmetry charges of a quantum phase of matter into a  
\emph{sandwich}, which is a topological order $\mathcal{T}$ defined on a slab with appropriate boundary conditions. One of the main benefits of this approach is to separate the topological data from the dynamical contents in the theory of interest. Fig.~\ref{fig:sandwich} illustrates the main idea. The bulk of the slab is a topological order that support topological gapped boundary conditions, which will be called a \emph{SymTFT}. The left boundary is chosen to be a topological gapped boundary of the topological order SymTFT, on which some bosonic anyons condensed. The confined topological defect lines on the left topological boundary encodes all the topological data of the symmetry in the bosonic systems. We will thus call the topological boundary on the left of the sandwich the \emph{symmetry boundary}. Fixing the symmetry boundary, different choices of the right boundary result in different bosonic quantum phases. We will refer the right boundary as the \emph{physical boundary}. We note that the right boundary could be non-topological, such as a CFT. Different phases can be analysed on the same footing only when we fix a choice of the symmetry boundary.

\paragraph{Sandwich}
Focusing the discussion on the \spd{1} bosonic systems, consider a \spd{2} SymTFT on an open 3D manifold $M\times I$, where $I=[0,1]$ is a 1D interval\footnote{In this work, we work in Euclidean signature.}, and $M$ is a closed surface. We take $M\times\{0\}$ as the symmetry boundary, and $M\times\{1\}$ as the physical boundary. The partition function on this \spd{2} manifold can be viewed as the inner product of two states:
\begin{equation}
    Z=\braket{\mathcal{A}|\Psi},
\end{equation}
where $\ket{\mathcal{A}}$ is the boundary state on the symmetry boundary, which is topological, and $\ket{\Psi}$ is the boundary state on the physical boundary. When the surface $M$ is not a sphere (i.e. nonzero genus), the Hilbert space of the bulk topological theory on $M$ has dimension greater than 1. There is a canonical choice for the basis states $\ket{\alpha}$. When $M$ is a torus, the label $\alpha$ corresponds to an anyon type in the bulk, and the state $\ket{\alpha}$ can be obtained as the basis of the Hilbert space on the solid torus $D^2\times S^1$, where $D^2$ is a disk, with an insertion of anyon $\alpha$ wrapping around $S^1$.

The physical boundary state $\ket{\Psi}$, which may or may not be topological, can always be expanded in the anyon basis $\ket{\alpha}$:
\begin{equation}
    \ket{\Psi} = \sum_{\alpha \in \mathcal{T}} Z_{\alpha} \ket{\alpha}.
\end{equation}
When $M$ is a torus $T^{2}$, the coefficient $Z_{\alpha}$ in this expansion is equal to the partition function defined on the solid torus $D^{2} \times S^{1}$ with $\Psi$ boundary condition and an insertion of anyon $\alpha$ around $S^{1}$.

Gapped boundaries of a \spd{2} topological order are classified by the Lagrangian algebra $\mathcal{A} = \bigoplus_{\alpha \in \mathcal{T}} w_{
\alpha} \alpha$, where $w_{\alpha}$ are some non-negative integers (see Appendix~\ref{app:superL} for a brief review). Each topological gapped boundary corresponds to a state 
\begin{equation}
    \ket{\mathcal{A}} = \sum_{\alpha \in \mathcal{T}} w_{
\alpha} \ket{\alpha}.
\end{equation}
Physically, the Lagrangian algebra describes condensation of anyons on the boundary: if $w_\alpha>0$, then an anyon $\alpha$ can condense on the boundary.

In the anyon basis, the partition function of the \spd{1} theory can be computed by taking the inner product:
\begin{align}
    Z &= \ip{\mathcal{A}}{\Psi} \nonumber
    \\
    &= \sum_{\alpha, \beta \in \mathcal{T}}w_{\beta} Z_{\alpha} \ip{\beta}{\alpha} \nonumber
    \\
    &=\sum_{\alpha \in \mathcal{T}}w_{\alpha} Z_{\alpha}.
\label{eq:sandz}
\end{align}
Therefore, the partition function of the \spd{1} theory can be written as a linear combination of the partition function of the SymTFT $\mathcal{T}$ on the solid torus with an insertion of of Lagrangian algebra $\mathcal{A}$ around $S^{1}$. The construction can also be generalized to higher genus surfaces.  

\paragraph{Symmetries and operators}
In the sandwich construction, the topological defect lines confined on the left symmetry boundary implement the symmetry of our system. In general, the topological defect lines are described by a fusion category $\mathcal{C}$. The anyons in the bulk SymTFT are then described by a modular tensor category called the Drinfeld center $\mathcal{Z}(\mathcal{C})$.

Local operators can be organized according to their transformations under the fusion category symmetry. As a result, a local operator can be uniquely attached a bulk anyon label $\alpha$ with $w_\alpha > 0$. Intuitively, the local operator $\mathcal{O}_{\alpha}$ corresponds to a bulk anyon line $\alpha$ that condenses on the left topological boundary, and stretches across the sandwich.

The symmetry $\mathcal{C}$ acts on the local operator as the \emph{half braiding} between the bulk anyon line $\alpha \in \mathcal{Z}(\mathcal{C})$ and the boundary topological line $m \in \mathcal{C}$ \cite{Hung2019,Lin2022}. For abelian SymTFT, it's usually convenient to move the boundary symmetry line into the bulk and the symmetry charges are encoded in the bulk anyon braiding phases. 

\paragraph{Phases}
Topological holography provides a unified framework to describe gapped and gapless phases. Once the symmetry boundary is fixed with a fusion category symmetry $\mathcal{C}$, different choices of the physical boundary realize different phases of matter. If the physical boundary is also a topological gapped boundary, the sandwich construction realizes a gapped phase in bosonic systems with the symmetry $\mathcal{C}$. If we choose the physical boundary to be gapless, such as a CFT, the sandwich construction realizes a critical point or a gapless phase with the fusion category symmetry $\mathcal{C}$. Such gapless boundaries are also characterized by condensable algebras of the form $\mathcal{B} = \bigoplus_{\beta \in \mathcal{T}} w_{
\beta} \beta$ \cite{Bhardwaj2024gspt}, which satisfy the conditions listed in Appendix~\ref{app:superL} except the Lagrangian condition. 

\paragraph{Dualities}
Topological holography provides a powerful way to systematically understand dualities between \spd{1} theories. Given a bulk SymTFT $\mathcal{T}$, in general there can be automorphisms that permutes the bulk anyons: 
\begin{equation}
    \rho: \alpha \rightarrow \rho(\alpha).
\end{equation}
Each of this automorphism is an anyonic symmetry of the SymTFT and corresponds to an invertible codimension-1 twist defect $\mathcal{D}$ in the bulk, such that a bulk anyon is permuted when it passing through the defect. In the sandwich construction, we can insert this twist defect parallel to the sandwich, giving rise to a new \spd{1} phase.

More specifically, consider two topological boundary states $\ket{\mathcal{L}_{1}}$ and $\ket{\mathcal{L}_{2}}$ on the physical boundary. The corresponding partition functions are given by $Z_{1} = \ip{\mathcal{A}}{\mathcal{L}_{1}}$ and $\z_{2} = \ip{\mathcal{A}}{\mathcal{L}_{2}}$, where $\bra{\mathcal{A}}$ is the topological boundary state on the symmetry boundary. We say that there is a duality between these two gapped phases, if there exist a twist defect $\mathcal{D}$, such that $\mathcal{D} \ket{\mathcal{L}_{1}} = \ket{\mathcal{L}_{2}}$. Written in terms of the partition functions, this means that $ Z_{2}= \ip{\mathcal{A}}{\mathcal{L}_{2}} = \me{\mathcal{A}}{\mathcal{D}}{\mathcal{L}_{1}}$. 

We can also consider dualities between two critical theories. For two critical boundaries $\Psi_{1}$ and $\Psi_{2}$ related by the twist defect of this kind: $\mathcal{D} \ket{\Psi_{1}} = \ket{\Psi_{2}}$, there is a duality between the partition functions of these two critical theories:
\begin{align}
     Z_{2} &= \ip{\mathcal{A}}{\Psi_{2}}
     \nonumber
     \\
     &= \me{\mathcal{A}}{\mathcal{D}}{\Psi_{1}} \nonumber
     \\
     &= \ip{\mathcal{A}'}{\Psi_{1}},
\label{eq:zrelation}
\end{align}
where $\bra{\mathcal{A}'} = \bra{\mathcal{A}} \mathcal{D}$. This shows that the partition function of theory $2$ is given by the sandwich of theory $1$ but with a different choice of the symmetry boundary. We can thus use this relation to write down the partition function for theory $2$ by using the characters of theory $1$. This will be the main strategy of this work. 

In general, even for any two topological gapped boundaries $\mathcal{A}$ and $\mathcal{B}$ not related by an anyonic symmetry, there is a corresponding duality between the two bosonic quantum systems, simply given by changing the topological gapped boundary condition on the symmetry boundary. This class of dualities is more general since the fusion category symmetries corresponding to the gapped boundaries $\mathcal{A}$ and $\mathcal{B}$ might not be equivalent in the sense of fusion categories. More formally, two quantum systems with fusion category symmetries $\mathcal{S}$ and $\mathcal{S}'$ are dual to each other in this generalized sense if $\mathcal{Z}(\mathcal{S}) \cong \mathcal{Z}(\mathcal{S}')$.

\section{Fermionization in topological holography}
\label{sec:fermionization}
To specify a theory of fermions on a general Riemann surface, it requires a choice of spin structure. The spin structure represents whether the fermions are periodic (P) or anti-periodic (A) around each cycle\footnote{Fermion being anti-periodic around a given cycle corresponds to the Neveu-Schwarz (NS) spin structure and periodic corresponds to the Ramond (R) spin structure in string theory.}. We denote this spin structure by $\rho$. 

Given a bosonic theory with a non-anomalous $\zz^{B}$ symmetry, a fermionic theory can be obtained by the fermionization procedure:
\begin{equation}
    Z_{F}(\rho) = \frac{1}{2^{g}} \sum_{a} Z_{B}(a) (-1)^{\text{Arf}(\rho+a)},
\label{eq:fermionization}
\end{equation}
where $\rho$ is the spin structure of the Riemann surface with genus $g$, $a$ is the background gauge field for the $\zz^{B}$ symmetry, and $\text{Arf}(\cdot)$ is the Arf invariant. This definition of the fermionization corresponds to the Jordan-Wigner transformation on the lattice. 

The inverse map is given by the bosonization:
\begin{equation}
    Z_{B}(a) = \frac{1}{2^{g}} \sum_{\rho} Z_{F}(\rho) (-1)^{\text{Arf}(\rho+a)}.
\end{equation}
The bosonization map however is not unique since there is a freedom of stacking a $(1+1)d$ Kitaev chain with partition function $(-1)^{\text{Arf}(\rho)}$, which produces anther fermion theory $\mathcal{T}_{F'}$. Performing the bosonization for $\mathcal{T}_{F'}$ givens another bosonic theory $\mathcal{T}_{B'}$:
\begin{equation}
    Z_{B'}(a) = \frac{1}{2^{g}} \sum_{\rho} Z_{F}(\rho) (-1)^{\text{Arf}(\rho+a) + \text{Arf}(\rho)}.
\label{eq:fermionization-arf}
\end{equation}
The two bosonic theories $\mathcal{T}_{B}$ and $\mathcal{T}_{B'}$ are related by gauging the $\zz^{B}$ symmetry, which corresponds to the Kramers-Wannier duality. The relations between the bosonic and fermionic theories are summarized in the diagram below:
\begin{equation}
\begin{tikzcd}[row sep=large, column sep=10em]
\mathcal{T}_{B} \arrow[d, leftrightarrow, "\text{B/F}"'] \arrow[r, leftrightarrow, "\zz^{B} \text{gauging}"]   & \mathcal{T}_{B'} \arrow[d, leftrightarrow, "\text{B/F}"]   
\\
\mathcal{T}_{F} \arrow[r, leftrightarrow, "\text{Stacking Kitaev}"]   & \mathcal{T}_{F'}
\end{tikzcd}
\end{equation}
where the vertical arrows represent the bosonization/fermionizaion procedure. 

\begin{figure}
	\centering
	\includegraphics[width=0.45\textwidth]{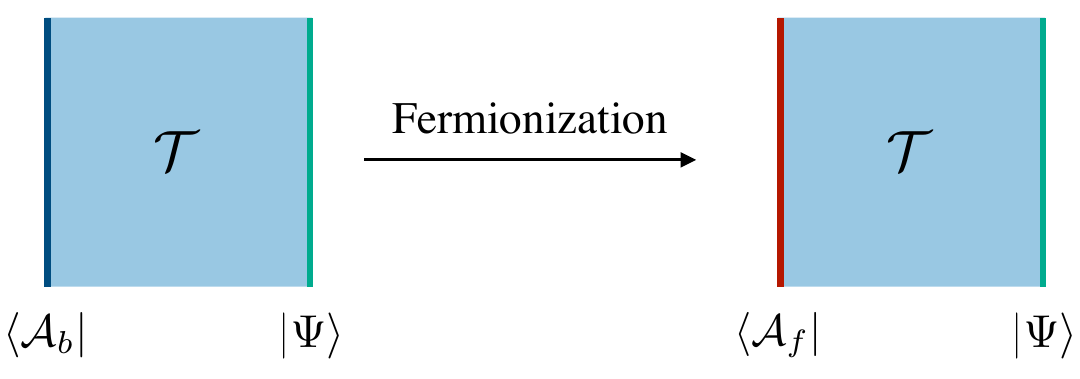}
	\caption{Performing the fermionization procedure in the sandwich picture corresponds to change the symmetry boundary from a bosonic gapped boundary $\mathcal{A}_{b}$ to a fermionic gapped boundary $\mathcal{A}_{f}$.}
	\label{fig:fsandwich}
\end{figure}

In general, performing the fermionization procedure in the sandwich picture corresponds to change the symmetry boundary from a bosonic gapped boundary to a fermionic one as discussed in Fig.~\ref{fig:fsandwich}~\cite{Gaiotto2016}. Fermionic gapped boundaries can be described as the fermion condensation and are classified by super Lagrangian algebra, which we review in Appendix.~\ref{app:superL}. The confined defect lines on the fermionic gapped boundary implement the symmetry in the fermionic systems, which in general is described by a super fusion category~\cite{Aasen2019,Lou2021} \footnote{We note that there is always a freedom to stack an invertible $(1+1)d$  Arf TQFT, which describes the Kitaev chian, to the fermionic gapped boundary and change the fermionic symmetry structure. See Ref.~\cite{Bhardwaj2024fsymtft} for more details. In this work, we don't consider this stacking.}.

Since we will be working in the anyon basis, it is useful to express the fermionization formula Eq.~\eqref{eq:fermionization} in terms of the multi-components partition functions in the anyon basis. On the torus, Eq.~\eqref{eq:fermionization} can be written as
\begin{equation}
    Z_{F}[m,n] = \frac{1}{2} \sum_{r,s} (-1)^{(m+r)(n+s)} Z_{B}[r,s],
\label{eq:fermionization-2}
\end{equation}
where $(m,n) \in \{ 0,1 \}$ labels the spin structure along the $x$- and $t$-direction on the torus, respectively, and $(r,s) \in \{ 0,1 \}$ labels the number of $\zz$ defects in the temporal and spatial directions, respectively. Here we use the convention that $m,n = 0$ is anti-periodic denoting as the ``A" structure and $m,n = 1$ is periodic denoting the ``P" structure. The twisted partition function can be written in the anyon basis~\cite{Huang2023symtft}:
\begin{align}
    Z[g,h] &= \me{W^{t}_{\mu_{g}}\mathcal{A}}{W^{x}_{\nu_{h}}}{\psi} \nonumber
    \\
    &= \sum_{\alpha, \gamma \in \mathcal{Z}(\mathcal{C})} w_{\alpha} N_{\mu_{g},\alpha}^{\gamma} \frac{S_{\gamma \nu_{h}}}{S_{0 \gamma}} Z_{\gamma},
\label{eq:ztwistanyon}
\end{align}
where $\mu_{g}$ and $\nu_{h}$ are anyons in the bulk SymTFT that implement the $g$ and $h$ symmetry \footnote{It's possible that $\mu_{g}$ and $\nu_{h}$ have non-trivial half-braiding phases as they are moved into the symmetry boundary~\cite{Bhardwaj2023}. In this work, we always choose the bulk representatives such that the half-braiding phases are trivial and the boundary lines can be freely pulled into the bulk without any phase factors.}. To perform the fermionization, we first identify $\mu_{f}$ to be the bulk anyon line that implement the $\zz$ symmetry participating in the fermionization and let $\mathcal{A}_{0}$ denotes the Lagrangian algebra of the bosonic system. Pluging Eq.~\eqref{eq:ztwistanyon} into Eq.~\eqref{eq:fermionization-2}, we obtain the fermionic partition function in the $(\text{A},\text{A})$ spin structure:
\begin{equation}
    Z_{F}[\text{A},\text{A}] = \frac{1}{2} \left( \sum_{\alpha \in A_{0}^{(+)}} w_{\alpha} Z_{\alpha} +  \sum_{\alpha \in (\mu_{f} \times A_{0})^{(-)}} w_{\alpha} Z_{\alpha} \right),
\label{eq:Znsns}
\end{equation}
where $A_{0}^{(+)}$ is the set of anyons in the Lagrangian algebra $\mathcal{A}_{0}$ that braid trivially with $\mu_{f}$, and $(\mu_{f} \times A_{0})^{(-)}$ is the set of anyons in $(\mu_{f} \times A_{0})$ that has a $(-1)$ mutaul braiding phase with the anyon $\mu_{f}$. Partition functions in the other sectors can be obtained similarly:
\begin{equation}
    Z_{F}[\text{P},\text{A}] = \frac{1}{2} \left( \sum_{\alpha \in A_{0}^{(-)}} w_{\alpha} Z_{\alpha} +  \sum_{\alpha \in (\mu_{f} \times A_{0})^{(+)}} w_{\alpha} Z_{\alpha} \right),
\label{eq:Zrns}
\end{equation}
\begin{equation}
    Z_{F}[\text{A},\text{P}] = \frac{1}{2} \left( \sum_{\alpha \in A_{0}^{(+)}} w_{\alpha} Z_{\alpha} -  \sum_{\alpha \in (\mu_{f} \times A_{0})^{(-)}} w_{\alpha} Z_{\alpha} \right),
\label{eq:Znsr}
\end{equation}
\begin{equation}
    Z_{F}[\text{P},\text{P}] = \frac{1}{2} \left( - \sum_{\alpha \in A_{0}^{(-)}} w_{\alpha} Z_{\alpha} +  \sum_{\alpha \in (\mu_{f} \times A_{0})^{(+)}} w_{\alpha} Z_{\alpha} \right).
\label{eq:Zrr}
\end{equation}
We have defined $A_{0}^{(-)}$ to be the set of anyons in the Lagrangian algebra $\mathcal{A}_{0}$ that has a $(-1)$ mutual braiding phase with $\mu_{f}$, and $(\mu_{f} \times A_{0})^{(+)}$ is the set of anyons in $(\mu_{f} \times A_{0})$ that braids trivially with $\mu_{f}$. 

We note that, in writing Eq.~\eqref{eq:Znsns}-\eqref{eq:Zrr}, a specific fermionic gapped boundary with the super Lagrangian algebra $A_{f}$ is chosen, where the anyons in the set $A_{0}^{(+)}$ and $(\mu_{f} \times A_{0})^{(-)}$ are condensed on the symmetry boundary. More formally, we find the set of anyons in the super Lagrangian algebra $A_{f} = A_{0}^{(+)} \cup (\mu_{f} \times A_{0})^{(-)}$ and the super Lagrangian algebra takes the form: 
\begin{equation}
    \mathcal{A}_{f} = \oplus_{\alpha \in A_{f}} w_{\alpha} \alpha,
\label{eqn:superlform}
\end{equation}
where $w_{\alpha}$ is determined by the Lagrangian algebra on the symmetry boundary before we perform the fermionization. 

In general there are multiple choices of the fermionic gapped boundary to perform the fermionization, which corresponds to fermionizing with respecting to different $\zz$ subgroup and the freedom of the stacking with an Kitaev chain. In this work, we will often choose the fermionic gapped boundary that corresponds to the Jordan–Wigner transformation on the lattice. 

\begin{figure}
	\centering
	\includegraphics[width=0.2\textwidth]{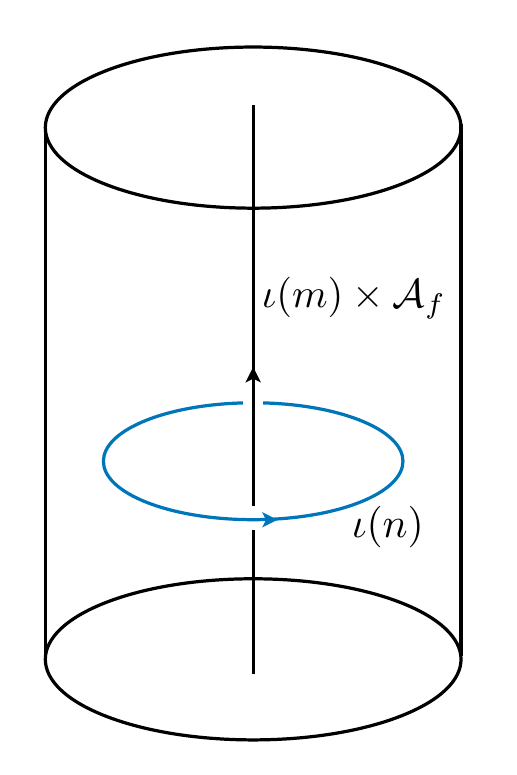}
	\caption{The fermionic partition functions with different spin structures are represented as the twisted partition functions $Z_{F}[m,n]$ in the anyon basis. We denote the representative anyon in the bulk SymTFT that implements the fermion parity symmetry as $\mu_\rho$ and define a map $\iota$ from the spin structures to the bulk anyons (see main text). The twisted partition function is equivalent to the partition function of the $(2+1)$D SymTFT on a solid torus with an insertion of a composite anyon $\iota(m) \times \mathcal{A}_{f}$ along the time-direction, wrapping around by $\iota(n)$ in the spatial direction. The temporal cycle is the vertical direction and the spatial cycle is horizontal.}
	\label{fig:twistz}
\end{figure}

Suppose we have fixed a fermionic gapped boundary condition $\mathcal{A}_{f}$ on the symmetry boundary of the sandwich and denote the corresponding boundary state as $\ket{\mathcal{A}_{f}}$. Let $\mu_\rho$ be the anyon line that generates the fermion parity symmetry in the bulk SymTFT \footnote{Again, we choose a bulk representative $\mu_\rho$ such that the half-braiding phase is trivial.}. The spin structure $(m,n)$ can be understood as the insertion of $m$ and $n$ numbers of the fermion parity line along the temporal and the spatial cycles, respectively. We define a map $\iota$ that maps the spin structures $(m,n)$ to the bulk anyons:
\begin{align}
    \iota: m &\rightarrow 1 \text{ , if } m=0
    \\
    m &\rightarrow \mu_{\rho} \text{ , if } m=1,
\end{align}
and similarly for $n$. The fermionic partition function with different spin structures are represented as the diagram shown in Fig.~\ref{fig:twistz}, and can be obtained directly by the following formula:
\begin{align}
    Z_{F}[m,n] &= \me{W_{\iota}\mathcal{A}_{f}}{V_{\iota}}{\psi} \nonumber
    \\
    &= \sum_{\alpha, \gamma \in \mathcal{Z}(\mathcal{C})} w_{\alpha} N_{\iota,\alpha}^{\gamma} \frac{S_{\gamma \iota}}{S_{0 \gamma}} Z_{\gamma},
\label{eq:ztwistanyon-f}
\end{align}
where $W_{\iota}$ and $V_{\iota}$ denotes the anyon line $\iota$ along $t$ and $x$-cycles, respectively.

\section{Fermionic chains with $\zz^{F}$ symmetry}
\label{sec:z2f}
Here we discuss the topological holographic picture for $(1+1)d$ fermionic chain with only $\zz^{F}$ symmetry and free Majorana fermion CFT. In the sandwich construction, we can always choose the bulk SymTFT to be described by a $\zz$ toric code. The symmetry boundary is chosen to be a fermionic gapped boundary condition. In the $\zz$ toric code, the fermionic gapped boundary is described by the super Lagrangian algebra:
\begin{equation}
    \mathcal{A}_{f} = 1 \oplus f,
\end{equation}
The confined topological defect line on the symmetry boundary is given by 
\begin{equation}
    X_{f} = e \oplus m,
\end{equation}
satisfying a $\zz$ fusion rule $X_{f} \otimes X_{f} = 1$. Physically, $X_{f}$ generates the fermion parity symmetry on the symmetry boundary. When $X_{f}$ line be pulled into the bulk SymTFT, it has a choice of being an $e$ or $m$ line. Here we choose $m$ line as the bulk representative of the $\zz^{F}$ fermion parity symmetry. The two choices are related by the EM exchange automorphism $e \leftrightarrow m$ and will lead to a duality between different gapped phases. 

\subsection{Gapped phases}
If we choose both symmetry and physical boundaries to be gapped boundary conditions, the sandwich construction realizes $(1+1)d$ gapped phases with $\zz^{F}$ fermion parity symmetry. There are only two possibilities: the trivial and the Kitaev phase. If the physical boundary is the $m$-condensed boundary, it corresponds to the trivial phase. This can be derived from the fact that the partition functions with different spin structures are all equal to $1$ by using Eq.~\eqref{eq:Znsns}-\eqref{eq:Zrr}. 

The Kitaev chain corresponds to choosing the physical boundary to be the $e$-condensed boundary. One can check that the partition function in the $(P,P)$ spin structure is given by
\begin{equation}
    Z[\text{P},\text{P}] = \frac{S_{em}}{S_{0e}} \ip{e}{e} = -1
\end{equation}
where we have used Eq.~\eqref{eq:ztwistanyon-f}. The other partition functions are all equal to $1$. Physically, this is the statement that the ground state of a Kitaev chain carries an odd fermion parity with a periodic boundary condition. Note that this identification of the phases is tied to the choice of the symmetry line of the $\zz^{F}$ fermion parity being the $m$ anyon line in the bulk SymTFT. If we choose $e$ to be the bulk representative of the fermion parity symmetry, the identification of the gapped phases will reverse. 

We note that our choice is consistent with the Jordan–Wigner transformation. The Jordan–Wigner transformation maps the trivial phase of an Ising spin chain to the trivial phase of the fermionic chain, and maps the $\zz$ SSB phase to the Kitave phase. In the sandwich picture of bosonic systems with a $\zz$ symmetry, we choose the $e$ condensed boundary as our symmetry boundary on the left. It follows that choosing the right boundary to be the $m$ condensed boundary produces the $\zz$ trivial phase and choosing the $e$ condensed boundary on the right produces the $\zz$ SSB phases. Upon performing the fermionization, we change the symmetry boundary on the left to be the $f$ condensed boundary and fix $m$ line to be the $\zz^{F}$ symmetry generator. From the above discussion, we see that the fermionization in the SymTFT framework indeed maps $\zz$ trivial phase to the trivial phase of the fermionic chain and maps the $\zz$ SSB phase to the Kitaev phase.

\subsection{Free Majorana fermion}
\label{sec:freemajorana}

The free Majorana fermion CFT can be obtained by fermionizing an Ising CFT. Upon performing the fermionization, we simply choose the symmetry boundary to be the $f$-condensed boundary condition, and the physical boundary is described by the Ising CFT. We denote the critical boundary state on the physical boundary as $\ket{\Psi_{\text{Ising}}}$. The untwisted partition function is given by the overlap:
\begin{align}
    Z[\text{A},\text{A}] &= \ip{\mathcal{A}_{f}}{\Psi_{\text{Ising}}} \nonumber
    \\
    &= Z_{1} + Z_{f} \nonumber
    \\
    &= |\chi_{1}|^{2} + |\chi_{\psi}|^{2} + \chi_{1} \overline{\chi}_{\psi} + \chi_{\psi} \overline{\chi}_{1},
\label{eq:freeZ}
\end{align}
where in the second equality we have used the multi-component partition functions of the Ising CFT in the $\zz$ toric code anyon basis \cite{Huang2023symtft}
\begin{align}
    Z_{1} &= |\chi_{1}|^{2} + |\chi_{\psi}|^{2}, \nonumber
    \\
    Z_{e} &= |\chi_{\sigma}|^{2}, \nonumber
    \\
    Z_{m} &= |\chi_{\sigma}|^{2}, \nonumber
    \\
    Z_{f} &= \chi_{1} \overline{\chi}_{\psi} + \chi_{\psi} \overline{\chi}_{1},
\label{eq:Ising_TC}
\end{align}
By comparing with Eq.~\eqref{eq:fermionization}, one can check that Eq.~\eqref{eq:freeZ} is indeed the fermion partition function of the free Majorana fermion with $(\text{A},\text{A})$ spin structure. 

Now we consider the fermionic partition functions twisted by the fermion parity. In the sandwich picture, they are represented by inserting $X_{f}$ line along $x$ and/or $t$ direction. Using Eq.~\eqref{eq:ztwistanyon}, we find
\begin{align}
    Z[\text{A},\text{P}] &= Z_{1} - Z_{f} \nonumber
    \\
    &= |\chi_{1}|^{2} + |\chi_{\psi}|^{2} - (\chi_{1} \overline{\chi}_{\psi} + \chi_{\psi} \overline{\chi}_{1}),
\\
     Z[\text{P},\text{A}] &= Z_{e} + Z_{m} \nonumber
     \\
     &= 2 |\chi_{\sigma}|^{2},
\\
     Z[\text{P},\text{P}] &= Z_{e} - Z_{m} \nonumber
     \\
     &= 0.
\end{align}
These are indeed the partition function for the free Majorana theory. 

Since all minimal models has a non-anomalous $\zz$ symmetry, all of them can be fermionized into \emph{fermionic minimal models}. The sandwich picture for the fermionic minimal models is discussed in Appendix~\ref{app:fminimal}.

\section{Fermionic chains with $\zz \times \zz^{F}$ symmetry}
\label{sec:z2z2f}

We now discuss the topological holographic picture fermionic chains with $\zz \times \zz^{F}$ symmetry. The bulk of the sandwich is a $\zz \times \zz$ gauge theory. For later convenience, we first review some properties of the $\zz \times \zz$ gauge theory and the corresponding sandwich picture for the bosonic cases. We sometimes denote the first $\zz$ by $\zz^{a}$ and the second one by $\zz^{b}$.

There are 6 gapped boundaries as summarized in Table~\ref{Table:z2z2bdy}. The automorphism in the $\mathbb{Z}_{2} \times \mathbb{Z}_{2}$ gauge theory are generated by the EM exchange for the $\mathbb{Z}_{2}^{a}$ and $\mathbb{Z}_{2}^{b}$ gauge theory:
\begin{equation}
    \sigma_{i} : e_{i} \leftrightarrow m_{i},
\end{equation}
and the twist defect from the gauged $\mathbb{Z}_{2} \times \mathbb{Z}_{2}$ SPT:
\begin{align}
    s: e_{i} \rightarrow e_{i}, \ m_{i} \rightarrow m_{i} e_{j}, i \neq j.
\label{eq:z2z2sdefect}
\end{align}

Choosing the symmetry boundary to be the $e$ condensed boundary $\mathcal{L}_{1}$ and physical boundary to be $\mathcal{L}_{i}$ for $i =1$ to $6$ realizes bosonic gapped phases with the unbroken subgroup and the SPT index $H^{2}(G,\U)$ shown in the second and third columns in Table.~\ref{Table:z2z2bdy}. The automorphisms correspond to the dualities between these gapped phases. 

\begin{table*}[t]\centering
	\begin{tabular}{c|c|c|c}
		Lagrangian algebra & Unbroken subgroup & $H^{2}(G,\U)$ & Order parameters
		\\
		\hline
		$\mathcal{L}_{1} = 1\oplus e_{1}\oplus e_{2}\oplus e_{1}e_{2}$ & $1$ & $1$ & $\mathcal{O}_{\mathbb{Z}_{2}^{a}} = W_{e_{1}}$, $\mathcal{O}_{\mathbb{Z}_{2}^{b}} = W_{e_{2}}$ 
		\\ 
		\hline
		$\mathcal{L}_{2} = 1\oplus m_{1}\oplus e_{2}\oplus m_{1}e_{2}$ & $\mathbb{Z}_{2}^{a}$ & 1 & $\mathcal{O}_{\mathbb{Z}_{2}^{b}} = W_{e_{2}}$
		\\ 
		\hline
		$\mathcal{L}_{3} = 1\oplus e_{1}\oplus m_{2}\oplus e_{1}m_{2}$ & $\mathbb{Z}_{2}^{b}$ & 1 & $\mathcal{O}_{\mathbb{Z}_{2}^{a}} = W_{e_{1}}$
		\\ 
		\hline
    	$\mathcal{L}_{4} = 1\oplus e_{1}e_{2}\oplus m_{1}m_{2}\oplus f_{1}f_{2}$ & $\mathbb{Z}_{2}^{{\rm diag}}$ & 1 & $\mathcal{O}_{\mathbb{Z}_{2}^{D}} = W_{e_{1}e_{2}}$
		\\ 
		\hline 
		$\mathcal{L}_{5} = 1\oplus e_{1}m_{2}\oplus  m_{1}e_{2}\oplus f_{1}f_{2}$ & $\mathbb{Z}_{2}^{a} \times \mathbb{Z}_{2}^{b}$ & $(-1)^{g_{1}h_{2}}$ & - 
		\\
		\hline
		$\mathcal{L}_{6} = 1\oplus m_{1}\oplus m_{2}\oplus m_{1}m_{2}$ & $\mathbb{Z}_{2}^{a} \times \mathbb{Z}_{2}^{b}$ & $1$ & -
		\\
		\hline
	\end{tabular}
	\caption{Gapped boundaries of the $\mathbb{Z}_{2} \times \mathbb{Z}_{2}$ gauge theory. The first column shows the Lagrangian algebra. The second and the third columns show the symmetry and the SPT index realized in the sandwich construction by choosing the symmetry boundary to be $\mathcal{L}_{1}$. The fourth column shows the order parameters as the Wilson line.}
	\label{Table:z2z2bdy}
\end{table*}

To obtain the topological holographic picture of the fermionic chains, we need to know the fermionic gapped boundary conditions of the $\zz \times \zz$ gauge theory, which are summarized in Table~\ref{Table:z2z2fbdy} \cite{Lou2021}. The fermionization is performed by choosing the symmetry boundary to be one of the fermionic gapped boundaries in Table~\ref{Table:z2z2fbdy}. Here we choose the symmetry boundary to be $\mathcal{L}^{f}_{1} = 1\oplus e_{1}e_{2}\oplus m_{1}f_{2}\oplus f_{1}m_{2}$. This choice corresponds to perform the fermionization with respecting to the diagonal $\zz$ symmetry:
\begin{equation}
    Z_{F}(\rho) = \frac{1}{2^{g}} \sum_{b} Z_{B}(b_{1}=b_{2}=b) (-1)^{\text{Arf}(\rho+b)},
\label{eq:fermionizationz2z2}
\end{equation}
where $b_{1}$ and $b_{2}$ are the $\zz$ gauge fields for the $\zz \times \zz$ symmetry. The fermion parity $\zz^{F}$ symmetry is implemented by the $m_{1}m_{2}$ line on the symmetry boundary and also the bulk and the $\zz$ symmetry is implemented by the $m_{1}$ line:
\begin{align}
    \zz^{F} &: m_{1}m_{2}, \nonumber
    \\
    \zz &: m_{1}.
\end{align}

\begin{table*}[t]\centering
	\begin{tabular}{c}
		Super Lagrangian algebra 
		\\
		\hline
		$\mathcal{L}^{f}_{1} = 1\oplus e_{1}e_{2}\oplus m_{1}f_{2}\oplus f_{1}m_{2}$
		\\ 
        \hline
        $\mathcal{L}^{f}_{2} = 1\oplus e_{1}f_{2}\oplus m_{1}m_{2}\oplus f_{1}m_{2}$
		\\ 
        \hline
        $\mathcal{L}^{f}_{3} = 1\oplus f_{1}\oplus f_{2}\oplus f_{1}f_{2}$
		\\
        \hline
        $\mathcal{L}^{f}_{4} = 1\oplus f_{1}\oplus e_{2}\oplus f_{1}e_{2}$
		\\ 
        \hline
        $\mathcal{L}^{f}_{5} = 1\oplus e_{1}\oplus f_{2}\oplus e_{1}f_{2}$
        \\ 
        \hline
        $\mathcal{L}^{f}_{6} = 1\oplus m_{1}\oplus f_{2}\oplus m_{1}f_{2}$
        \\ 
        \hline
		$\mathcal{L}^{f}_{7} = 1\oplus f_{1}\oplus m_{2}\oplus f_{1}m_{2}$
        \\ 
        \hline
		$\mathcal{L}^{f}_{8} = 1\oplus e_{1}f_{2}\oplus m_{1}e_{2}\oplus f_{1}m_{2}$
        \\ 
        \hline
        $\mathcal{L}^{f}_{9} = 1\oplus f_{1}e_{2}\oplus e_{1}m_{2}\oplus m_{1}f_{2}$
        \\ 
        \hline
	\end{tabular}
	\caption{Fermionic gapped boundaries of the $\mathbb{Z}_{2} \times \mathbb{Z}_{2}$ gauge theory described by the super Lagrangian algebra.}
	\label{Table:z2z2fbdy}
\end{table*}

\subsection{Gapped phases}
By choosing the physical boundary to be one of the bosonic gapped boundaries in Table.~\ref{Table:z2z2bdy}, the sandwich construction realizes fermionic gapped phases with $\zz \times \zz^{F}$ symmetry, which are summarized in Table~\ref{Table:z2z2fgapped}. 

The identification of the fermionic gapped phases is achieved by calculating the partition functions by using Eq.~\eqref{eq:ztwistanyon} and Eq.~\eqref{eq:ztwistanyon-f}. As a demonstration, let's focus on the case where the physical boundary is given by $\mathcal{L}_{2}$ and calculate the fermionic partition functions. The partition function in the $(\text{P},\text{P})$ structure without any $\zz$ twist is given by the diagram Fig.~\ref{fig:twistz} with $\rho = m_{1}m_{2}$:
\begin{equation}
    Z[\text{P},\text{P}] = \me{m_{1}m_{2}\times \mathcal{L}^{f}_{1}}{V^{t}_{m_{1}m_{2}}}{\mathcal{L}_{2}}
    = -1.
\end{equation}
One can check that the partition functions in the $(\text{P},\text{P})$ structure equal to $-1$ in any of the $\zz$ twist sectors. All other twisted partition functions equal to $1$. These are precisely the behavior of the partition functions of the Kitaev phase. We thus conclude that choosing the physical boundary to be $\mathcal{L}_{2}$ realizes the Kitaev phase. The same conclusion can be achieved by using the fermionization formula Eq.~\eqref{eq:fermionization} directly. We thus use $Z[\text{P},\text{P}]=-1$ to identify the phases that coexist with the Kitaev phase, which happens for $\mathcal{L}_{1}$ and $\mathcal{L}_{3}$ boundary. 

For phases that break the $\zz$ symmetry spontaneously, the prefactor in front of the partition functions is $2$, indicating the two-fold ground state degeneracy. For example, for the $\mathcal{L}_{4}$ boundary, we find that $Z[\text{A},\text{A}]= \ip{\mathcal{L}_{1}}{\mathcal{L}_{4}} = 2$ and all other twisted partition functions equal to $2$. Therefore, $\mathcal{L}_{4}$ corresponds to the $\zz$ SSB phase. We can use the absolute value of the partition functions to identify any phases that could coexist with the $\zz$ SSB phases, which happens for $\mathcal{L}_{1}$ boundary. We therefore find that the $\mathcal{L}_{1}$ boundary corresponds to a gapped phase that breaks the $\zz$ spontaneously with the remaining fermionic sector in the Kitaev phase. We denote this phase as ``SSB+Kitaev". 

The Gu-Wen fermionic SPT phase with $\zz \times \zz^{F}$ symmetry~\cite{Gaiotto2016} is characterized by the partition function on the torus:
\begin{equation}
    (-1)^{\int_{X} q_{\rho}(a)},
\label{eqn:gwz}
\end{equation}
where $\rho$ is the spin structure, $a \in H^{1}(X,\zz)$ is the $\zz$ gauge field, and $\int_{X} q_{\rho}(a)$ is a $\zz$-valued quadratic function of $a$ that satisfies 
\begin{equation}
    \int_{X} q_{\rho}(a + a') = \int_{X} q_{\rho}(a) + \int_{X} q_{\rho}(a') + a \cup a'.
\end{equation}
We can write the GW partition function Eq.~\eqref{eqn:gwz} more explicitly in terms of the quadratic form $p_{\rho}: H_{1}(X,\zz) \rightarrow \zz$ that satisfies 
\begin{align}
     p_{\rho}(\gamma_{1} + \gamma_{2}) = p_{\rho}(\gamma_{1}) + p_{\rho}(\gamma_{2}) + \gamma_{1} \cdot \gamma_{2},
\end{align}
$\gamma_{1}, \gamma_{2} \in H_{1}(X,\zz)$ and $\gamma_{1} \cdot \gamma_{2}$ is the intersection number. In particular,
\begin{align}
     p_{\rho}(\gamma) = 0
\end{align}
if the spin structure is anti-periodic around a non-trivial cycle $\gamma$,
\begin{align}
     p_{\rho}(\gamma) = 1
\end{align}
if the spin structure is periodic around a non-trivial $\gamma$, and $p_{\rho}(\gamma)=0$ if $\gamma$ is homologically trivial. The GW partition function can then be written as
\begin{equation}
    (-1)^{\int_{X} q_{\rho}(a)} = (-1)^{a(\gamma_{1})a(\gamma_{2})+p_{\rho}(\gamma_{1})a(\gamma_{2}) + a(\gamma_{1})p_{\rho}(\gamma_{2})},
\label{eqn:gwzcycle}
\end{equation}
where now $\gamma_{1}$ and $\gamma_{2}$ are the two non-trivial cycle on the torus. We denote the twisted partition functions as $Z[(\rho_{x},\rho_{t});(g_{x},g_{t})]$ where $(\rho_{x},\rho_{t})$ are the spin structure in the spatial and temporal directions and $(g_{x},g_{t})$ are the spatial and temporal $\zz$ twists. Using Eq.~\eqref{eqn:gwzcycle}, one can see that 
\begin{align}
    &Z[(A,A);(1,1)] = Z[(A,P);(1,0)]  \nonumber
    \\
    &= Z[(P,A);(0,1)] = Z[(P,P);(0,1)] \nonumber
    \\
    &= Z[(P,P);(1,0)] = Z[(P,P);(1,1)] = -1,
\label{eqn:gwztwist}
\end{align}
and the other partition functions are equal to $1$. We thus use Eq.~\eqref{eqn:gwztwist} to identify the GW phase, which happens for $\mathcal{L}_{3}$ and $\mathcal{L}_{5}$ boundaries. The results are summarized in the third column in Table~\ref{Table:z2z2fgapped}.

\begin{table*}[t]\centering
	\begin{tabular}{c|c|c}
		Lagrangian algebra & Bosonic partition function & Fermionic gapped phases
		\\
		\hline
		$\mathcal{L}_{1} = 1\oplus e_{1}\oplus e_{2}\oplus e_{1}e_{2}$ & $4\delta(A_{1})\delta(A_{2})$ & SSB + Kitave 
		\\ 
		\hline
		$\mathcal{L}_{2} = 1\oplus m_{1}\oplus e_{2}\oplus m_{1}e_{2}$ & $2\delta(A_{2})$ & Kitaev
		\\ 
		\hline
		$\mathcal{L}_{3} = 1\oplus e_{1}\oplus m_{2}\oplus e_{1}m_{2}$ & $2\delta(A_{1})$ & Kitaev+GW
		\\ 
		\hline
    	$\mathcal{L}_{4} = 1\oplus e_{1}e_{2}\oplus m_{1}m_{2}\oplus f_{1}f_{2}$ & $2\delta(A_{1}+A_{2})$ & SSB
		\\ 
		\hline 
		$\mathcal{L}_{5} = 1\oplus e_{1}m_{2}\oplus  m_{1}e_{2}\oplus f_{1}f_{2}$ & $(-1)^{A_{1}\cup A_{2}}$ & GW
		\\
		\hline
		$\mathcal{L}_{6} = 1\oplus m_{1}\oplus m_{2}\oplus m_{1}m_{2}$ & 1 & Trivial
		\\
		\hline
	\end{tabular}
	\caption{Fermionic gapped phases with $\zz \times \zz^{F}$ symmetry in topological holography. The left symmetry boundary is chosen to be the fermionic gapped boundary $\mathcal{L}^{f}_{1}$. The first column shows the physical boundary on the right of the sandwich. The second column shows that bosonic TQFT partition function of the bosonic gapped phases before the fermionization. The third column shows the fermoinic gapped phases realized by the sandwich construction. ``SSB" means that the $\zz$ symmetry is broken spontaneously. ``Kitaev" denotes the fermionic chain is in the Kitaev phase. ``GW" denotes the ``Gu-Wen" phase, which is the non-trivial fermionic SPT phase with $\zz \times \zz^{F}$ symmetry. ``Trivial" means the fermoinic chain is in the completely trivial symmetric phase. The fermoinic gapped phases obtained by the sandwich picture match with the fermionization results obtained directly by applying Eq.~\eqref{eq:fermionization}.
    }
	\label{Table:z2z2fgapped}
\end{table*}

\subsection{Quantum critical points}
\label{sec:z2z2fqcp}

\begin{figure}
	\centering
	\includegraphics[width=0.48\textwidth]{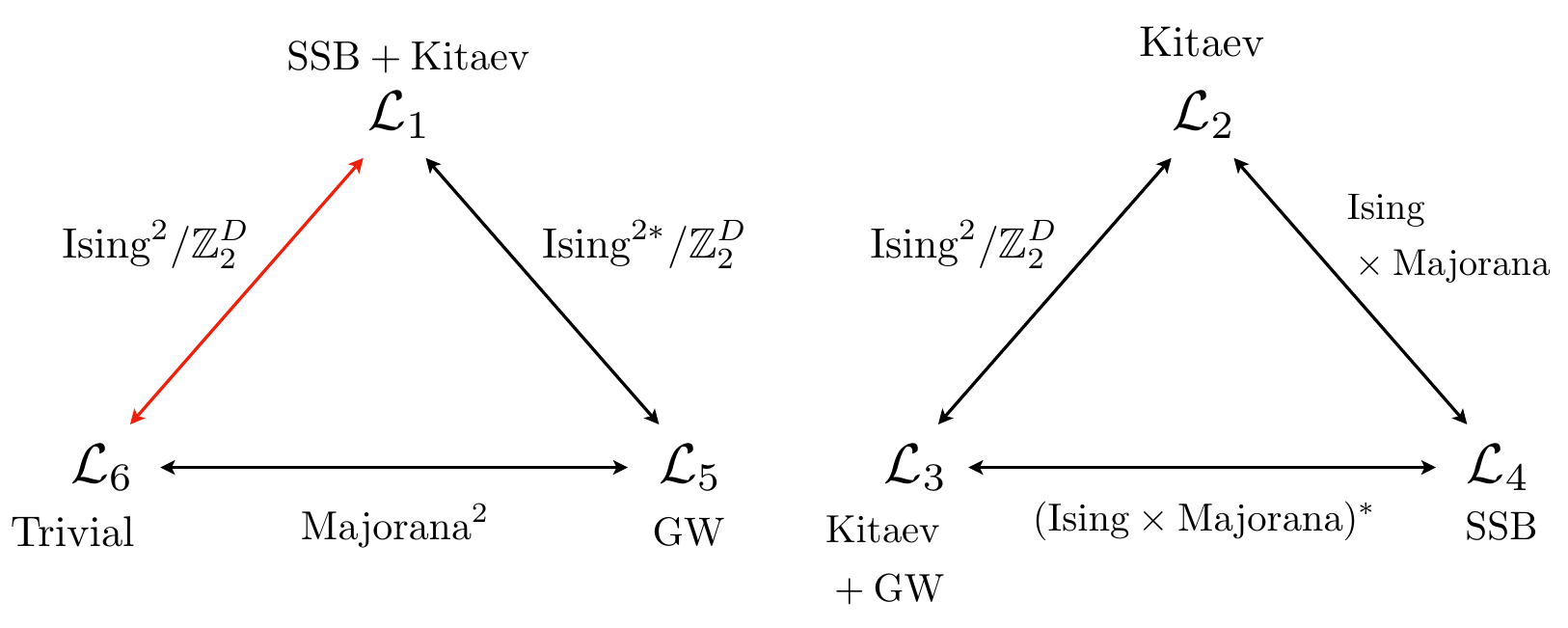}
	\caption{Summary of the phase transitions in the fermionic systems with $\zz \times \z_{2}^{F}$ symmetry discussed in Sec.~\ref{sec:z2z2fqcp} and Sec.~\ref{sec:fseqcp}. The red arrow indicate the input boundary critical theory on the physical boundary of the sandwich, which is the $\text{Ising}^{2}$ CFT.}
\label{fig:z2z2fftrans}
\end{figure}

We now discuss various quantum critical points in fermionic systems with $\zz \times \zz^{F}$ symmetry by using the sandwich picture. As we will see below, the inputs that we require are the CFT description of boundary phase transitions on the physical boundary. The input CFT that we focus on heavily in this and the next section is the $\text{Ising}^{2}$ CFT that describes the boundary critical point between the topological gapped boundary $\mathcal{L}_{1}$ and $\mathcal{L}_{6}$.

Some of the critical points in the $\zz \times \zz^{F}$ fermionic systems have non-trivial topological responses in the twisted sectors. Such critical points will be referred to as the symmetry enriched quantum critical points and will be discussed in Sec.~\ref{sec:fseqcp}. The results are summarized in Fig.~\ref{fig:z2z2fftrans}.

\subsubsection{$\text{Trivial} \leftrightarrow (\text{SSB} + \text{Kitaev}) = \text{Ising}^{2}  \bf{/Z_{2}^{D}} $}
We consider the transition between the trivial and the ``SSB + Kitaev" phase, where the $\zz$ symmetry is spontaneously broken during the transition. The trivial phase is realized by choosing the physical boundary to be $\mathcal{L}_{1}$ and the ``SSB + Kitaev" phase corresponds to $\mathcal{L}_{6}$. 

We now discuss the input boundary critical theory. It is helpful to consider the bosonic side first. From Table.~\ref{Table:z2z2bdy}, we see that $\mathcal{L}_{1}$ and $\mathcal{L}_{6}$ corresponds to complete SSB phase and the trivial symmetric phase respectively if we choose the symmetry boundary to be $\mathcal{L}_{1}$. If we tune the right boundary of the sandwich to be at the critical point between the gapped boundaries $\mathcal{L}_{1}$ and $\mathcal{L}_{6}$, we know that this critical point is described by the $\text{Ising}^{2}$ CFT. The CFT boundary state of the sandwich is then given by
\begin{equation}
    \ket{\psi_{16}} = \sum_{\alpha,\beta \in \text{TC}} Z_{\alpha} Z_{\beta} \ket{\alpha \beta},
\end{equation}
where $Z_{a}$ and $Z_{b}$ are the partition functions of the Ising CFT in the anyon sector $a$ and $b$. This is our input boundary critical theory on the physical boundary of the sandwich. Taking the overlap $\ip{\mathcal{L}_{1}}{\psi_{16}}$ produces the partition function for the $\text{Ising}^{2}$ CFT as expected. 

We now fermionize the theory by changing the symmetry boundary to be $\mathcal{L}_{1}^{f}$ and calculate the partition function as the overlap:
\begin{widetext}
\begin{align}
   Z^{\text{Ising}^{2}\boldsymbol{/Z_{2}^{D}}}[\text{A},\text{A}] &= \ip{\mathcal{L}_{1}^{f}}{\psi_{16}}  \nonumber
   \\
   &= Z_{1}^{2} + Z_{e}^{2} + 2 Z_{m}Z_{f}  \nonumber
   \\
   &= ( |\chi_{1}|^{2} + |\chi_{\psi}|^{2})^{2} + 
   2 |\chi_{\sigma}|^{2} ( \chi_{1} \overline{\chi}_{\psi} + \chi_{\psi} \overline{\chi}_{1}) + 4 |\chi_{\sigma}|^{2}.
\label{eqn:M2z2}
\end{align}
\end{widetext}

We now show that Eq.~\eqref{eqn:M2z2} is the fermionization with respecting to the diagonal $\zz$ symmetry. We rewrite Eq.~\eqref{eqn:M2z2} as 
\begin{widetext}
\begin{align}
   Z^{\text{Ising}^{2}\boldsymbol{/Z_{2}^{D}}}[\text{A},\text{A}] &= ( |\chi_{1}|^{2} + |\chi_{\psi}|^{2})^{2} + 
   2 |\chi_{\sigma}|^{2} ( \chi_{1} \overline{\chi}_{\psi} + \chi_{\psi} \overline{\chi}_{1}) + 4 |\chi_{\sigma}|^{2} \nonumber
   \\
   &= \frac{1}{2} \left( Z_{0,0}^{2} +  Z_{0,1}^{2} + Z_{1,0}^{2} - Z_{1,1}^{2} \right) \nonumber
   \\
   &= \frac{1}{2} \sum_{a} Z(a_{1} = a, a_{2} = a) (-1)^{\text{Arf}(\rho+a)},
\end{align}
\end{widetext}
where, in the second equality, we have used the twisted partition functions for the Ising CFT:
\begin{align}
    Z_{0,1} &= |\chi_{1}|^{2} + |\chi_{\psi}|^{2} - |\chi_{\sigma}|^{2},
    \\
    Z_{1,0} &= |\chi_{\sigma}|^{2} +  \chi_{1} \overline{\chi}_{\psi} + \chi_{\psi}\overline{\chi}_{1},
    \label{eqn:zising10}
    \\
     Z_{1,1} &= |\chi_{\sigma}|^{2} -  \chi_{1} \overline{\chi}_{\psi} - \chi_{\psi}\overline{\chi}_{1}.
\end{align}
Therefore, we see that the theory given by Eq.~\eqref{eqn:M2z2} can be obtained by performing the fermionization with respect to the diagonal $\zz$ symmetry. We denote this theory as $\text{Ising}^{2}\boldsymbol{/Z_{2}^{D}}$\footnote{We use the symbol ``$\boldsymbol{/Z_{2}}$" to denote fermionizing with respecting to a $\zz$ symmetry to distinguish with gauging a $\zz$ symmetry.}.

\subsubsection{$\text{Kitaev} \leftrightarrow (\text{Kitaev} + \text{GW}) = \text{Ising}^{2} \bf{/Z_{2}^{D}} $}

We proceed to discuss the transition between the Kitaev and the ``Kitaev + GW" phase. The Kitaev phase is realized by choosing the physical boundary to be $\mathcal{L}_{2}$ and the ```Kitaev + GW" phase corresponds to $\mathcal{L}_{3}$. We denote the corresponding state as $\ket{\psi_{2,3}}$ and the partition function of the SPT transition is $Z[{\text{A},\text{A}}] = \ip{\mathcal{L}_{1}^{f}}{\psi_{2,3}}$. We would like to relate the partition function of this transition to the input boundary transition $\ket{\psi_{1,6}}$. Such duality transformation is implemented by the insertion of the $D_{\sigma_{2}}$ twist defect. The partition function can thus be written as 
\begin{align}
    Z[\text{A},\text{A}] &= \ip{\mathcal{L}_{1}^{f}}{\psi_{2,3}}  \nonumber
    \\
    &= \me{\mathcal{L}_{1}^{f}}{\mathcal{D}_{2}}{\psi_{1,6}} \nonumber
    \\
    &= \ip{\mathcal{L}_{2}^{f}}{\psi_{1,6}} \nonumber
    \\
    &= Z_{1}^{2} + Z_{e}Z_{m} + Z_{m}Z_{f} + Z_{e}Z_{f} \nonumber
    \\
    &=Z^{\text{Ising}^{2}\boldsymbol{/Z_{2}^{D}}}[\text{A},\text{A}],
\label{eq:tgwz}
\end{align}
where $Z^{\text{Ising}^{2}\boldsymbol{/Z_{2}^{D}}}[\text{A},\text{A}]$ is given in Eq.~\eqref{eqn:M2z2}. We find that the transition is also described by the $\text{Ising}^{2}\boldsymbol{/Z_{2}^{D}}$ theory.

\subsubsection{$ \text{Trivial} \leftrightarrow \text{GW} = \text{Majorana}^{2} $}
Here we consider the transition between the trivial and the Gu-Wen SPT phases. This critical point is realized by tuning the physical boundary of the sandwich to be at the critical point between the gapped boundaries $\mathcal{L}_{5}$ and $\mathcal{L}_{6}$, while choosing the symmetry boundary to be the fermionic gapped boundary $\mathcal{L}_{1}^{f}$. We denote the corresponding state as $\ket{\psi_{5,6}}$ and the partition function of the SPT transition is $Z[{\text{A},\text{A}}] = \ip{\mathcal{L}_{1}^{f}}{\psi_{5,6}}$. We would like to obtain the partition function of this transition by applying some duality transformation. This is achieve by considering the twist defect $\mathcal{D} = \mathcal{D}_{\sigma_{1}\sigma_{2}} \circ \mathcal{D}_{s}$. The twist defect $\mathcal{D}$ acts on the gapped boundaries as 
\begin{equation}
    \mathcal{D} \ket{\mathcal{L}_{1}} = \ket{\mathcal{L}_{6}}, \ \mathcal{D} \ket{\mathcal{L}_{6}} = \ket{\mathcal{L}_{5}}.
\end{equation}
Therefore, we have
\begin{equation}
    \ket{\psi_{5,6}} = \mathcal{D} \ket{\psi_{1,6}}.
\end{equation}
The partition function for the transition is given by inserting the twist defect $\mathcal{D}$:
\begin{align}
    Z[\text{A},\text{A}] &= \ip{\mathcal{L}_{1}^{f}}{\psi_{5,6}}  \nonumber
    \\
    &= \me{\mathcal{L}_{1}^{f}}{\mathcal{D}}{\psi_{1,6}} \nonumber
    \\
    &= \ip{\mathcal{L}_{3}^{f}}{\psi_{1,6}} \nonumber
    \\
    &= Z_{1}^{2} + 2 Z_{1}Z_{f} + Z_{f}^{2} \nonumber
    \\
    &= (|\chi_{1}|^{2} + |\chi_{\psi}|^{2} + \chi_{1} \overline{\chi}_{\psi} + \chi_{\psi} \overline{\chi}_{1})^{2},
\label{eq:tgwz2}
\end{align}
where, in the third equality, we have used the fact that $\bra{\mathcal{L}_{1}^{f}}\mathcal{D} = \bra{\mathcal{L}_{3}^{f}}$. Eq.~\eqref{eq:tgwz2} is the partition function of two copies of free Majorana theories in the $(\text{A},\text{A})$ spin structure, which is equivalent to a Dirac theory. 

\subsubsection{$ \text{Kitaev} \leftrightarrow \text{SSB} = \text{Ising} \times \text{Majorana} $}
We consider the transition between the Kitaev and the $\zz$ SSB phases in this section, which corresponds to the boundary $\mathcal{L}_{2}$ and $\mathcal{L}_{4}$, respectively. We denote the corresponding critical boundary state as $\ket{\psi_{2,4}}$ and the partition function of the SPT transition is $Z[{\text{A},\text{A}}] = \ip{\mathcal{L}_{1}^{f}}{\psi_{2,4}}$. We would like to relate the critical boundary state $\ket{\psi_{2,4}}$ to $\ket{\psi_{1,6}}$, which we know is described by the $\text{Ising}^{2}$ CFT. To perform the duality, we consider the defect $\mathcal{D}_{\sigma_{1}} \circ \mathcal{D}_{s}$ such that the action is given by $\mathcal{D}_{\sigma_{1}} \circ \mathcal{D}_{s} \ket{\psi_{1,6}} = \ket{\psi_{2,4}}$. 
The partition function for the transition is given by
\begin{widetext}
\begin{align}
    Z[\text{A},\text{A}] &= \ip{\mathcal{L}_{1}^{f}}{\psi_{2,4}}  \nonumber
    \\
    &= \me{\mathcal{L}_{1}^{f}}{\mathcal{D}_{\sigma_{1}} \circ \mathcal{D}_{s}}{\psi_{1,6}} \nonumber
    \\
    &= \ip{\mathcal{L}_{6}^{f}}{\psi_{1,6}} \nonumber
    \\
    &= (Z_{1} + Z_{m})(Z_{1} + Z_{f}) \nonumber
    \\
    &=  (|\chi_{1}|^{2} + |\chi_{\sigma}|^{2} + |\chi_{\psi}|^{2}) (|\chi_{1}|^{2} + |\chi_{\psi}|^{2} + \chi_{1} \overline{\chi}_{\psi} + \chi_{\psi} \overline{\chi}_{1}),
\label{eq:tgwz3}
\end{align}
\end{widetext}
which is the stacking of the Ising CFT and a single Majorana theory.

\subsubsection{$ \text{Trivial} \leftrightarrow \text{Kitaev} = \text{Majorana}$}

Thus far, the transitions discussed in this section are based on the input boundary transition between $\mathcal{L}_{1}$ and $\mathcal{L}_{6}$, which is the $\text{Ising}^{2}$ CFT. Here we demonstrate that it's possible to obtain the other phase transitions by using different input boundary transitions.

We consider the transition between the Trivial and the Kitaev phases. From Table.~\ref{Table:z2z2fgapped}, we see that this transition sits between the gapped boundary $\mathcal{L}_{2}$ and $\mathcal{L}_{6}$. We now discuss a possible boundary phase critical theory as our input. We note that $\mathcal{L}_{2}$ and $\mathcal{L}_{6}$, on the bosonic side, corresponds to partial SSB phase (break the second $\zz$) and the trivial $\zz \times \zz$ symmetric phase respectively if we choose the symmetry boundary to be $\mathcal{L}_{1}$. Since only the second $\zz$ symmetry is broken, we can view the system as two layers of toric code sandwich. The transition described by the Ising CFT only happens at the physical boundary for second layer, and the first layer stays in the m-condensed gapped boundary. The input CFT boundary state on the physical boundary of the sandwich then takes the following form:
\begin{equation}
    \ket{\psi_{26}} = \sum_{\alpha,\beta \in \text{TC}} v_{\alpha} Z_{\beta} \ket{\alpha \beta},
\end{equation}
where TC is the set of anyons in the toric code (we always order the anyons as $(1,e,m,f)$), $v_{\alpha} = (1,0,1,0)$ describes the condensed anyons ($1$ and $m$) on the physical boundary for the first copy, $Z_{\beta}$ is the Ising partition function in the anyon sector $\beta \in $ TC for the second copy. If we calculate the partition function for the bosonic system as the overlap $Z = \ip{\mathcal{L}_{1}}{\psi_{26}}$, it gives the partition function of the Ising CFT.

Now we perform the fermionization by choosing the symmetry boundary to be $\mathcal{L}_{1}^{f}$ and calculate the overlap:
\begin{align}
   Z[\text{A},\text{A}] &= \ip{\mathcal{L}_{1}^{f}}{\psi_{26}}  \nonumber
   \\
   &= v_{1}Z_{1} + v_{e}Z_{e} + v_{m}Z_{f} + v_{f}Z_{m}  \nonumber
   \\
   &= Z_{1} + Z_{f}  \nonumber
   \\
   &= |\chi_{1}|^{2} + |\chi_{\psi}|^{2} + \chi_{1} \overline{\chi}_{\psi} + \chi_{\psi} \overline{\chi}_{1},
\end{align}
where we have used $v_{\alpha} = (1,0,1,0)$ in the third equality. This is precisely the partition function of the free Majorana theory.  

\section{Fermionic symmetry enriched quantum critical points}
\label{sec:fseqcp}

\begin{figure}
	\centering
	\includegraphics[width=0.45\textwidth]{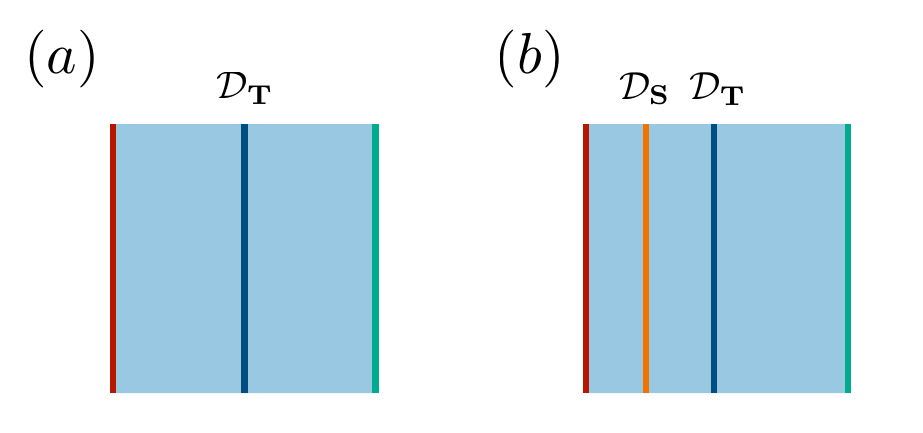}
	\caption{The sandwich pictures for two different types of the fermionic SEQCPs. (a) The first type of fermionic SEQCPs is obtain from the stacking of the usual SSB transitions with a SPT state. It corresponds to an insertion of a SPT twist defect $\mathcal{D}_{\textbf{T}}$ in the sandwich picture with the symmetry boundary that can ``absorb" the SPT defect. (b) The second type of fermionic SEQCPs is obtain from the a twisted gauging $\textbf{TS}$. It corresponds to an insertion of a twist defect $\mathcal{D}_{\textbf{T}} \circ \mathcal{D}_{\textbf{S}}$, where $\mathcal{D}_{\textbf{S}}$ is a defect that implements the $\textbf{S}$ gauging.}
\label{fig:twistgauging}
\end{figure}

Fermionic symmetry enriched quantum critical points (SEQCPs) are quantum critical points generally happens at the transition between a SSB phase, where the bosonic symmetry is completely broken, and a fermionic SPT phase. An important property of these quantum critical points is that the untwisted partition functions are the same as the partition function of the transition between the SSB and the trivial symmetric phases. However, the twisted partition functions differ from the usual SSB transitions and encode the SPT invariants.

We found that there are two different types of the fermionic SEQCPs. The first type of fermionic SEQCPs can be understood as the stacking of the usual SSB transitions with a SPT state (an operation we label as $\textbf{T}$). These kind of the SEQCPs are also called the gapless SPT phases~\cite{Scaffidi2017,verresen2021,zheng2022gaplessspt,Yu2022,Wen2023gspt}. In general, we expect the following form of the partition function:
\begin{align}
     Z^{\text{SEQCP}}[g,h] = Z^{\text{SSB}}[g,h] Z^{\text{SPT}}[g,h],
\end{align}
where $Z^{\text{SPT}}[g,h]$ is the SPT invariant and $Z^{\text{SSB}}[g,h]$ is the partition function of the symmetry breaking transition. In the sandwich picture, it corresponds to a SymTFT with an insertion of a SPT twist defect $\mathcal{D}_{\textbf{T}}$ and, in particular, the symmetry boundary can ``absorb" the defect $\mathcal{D}_{\textbf{T}}$, as illustrated in Fig.~\ref{fig:twistgauging}(a). 

The other type of the fermionic SEQCPs is obtained from the usual SSB transition by a twisted gauging $\textbf{TS}$, where $\textbf{S}$ is understood as gauging a finite subgroup in the fermionic symmetry group $G_{f}$. In the sandwich picture, it corresponds to an insertion of a twist defect $\mathcal{D}_{\textbf{TS}} = \mathcal{D}_{\textbf{T}} \circ \mathcal{D}_{\textbf{S}}$, where $\mathcal{D}_{\textbf{S}}$ is a defect that implements the $\textbf{S}$ gauging and $\mathcal{D}_{\textbf{T}}$ is a SPT twist defect, as shown in Fig.~\ref{fig:twistgauging}(b). We will present examples of both types of the fermionic SEQCPs in the $\zz \times \zz^{F}$ fermoinic systems in the following sections.

\subsection{$\text{GW} \leftrightarrow (\text{SSB} + \text{Kitaev}) = \text{Ising}^{2*} \bf{/Z_{2}^{D}}$}

Our first example of the symmetry enriched quantum critical point is the transition between the Gu-Wen and the ``SSB+Kitaev" phases. We tune the physical boundary of the sandwich to be at the critical point between the gapped boundaries $\mathcal{L}_{1}$ and $\mathcal{L}_{5}$. We denote the corresponding boundary state as $\ket{\psi_{1,5}}$ and the partition function of the transition is given by the overlap:
\begin{equation}
    Z^{\text{Ising}^{2*}\boldsymbol{/Z_{2}^{D}}}[\text{A},\text{A}] = \ip{\mathcal{L}_{1}^{f}}{\psi_{1,5}}.
\end{equation}
Now consider a twist defect $\mathcal{D}_{s}$ obtained from gauging the non-trivial $\zz \times \zz$ SPT state, which implements the anyon permutations shown in Eq.~\eqref{eq:z2z2sdefect}. The untwisted partition function (in the (A,A) spin structure) of the symmetry enriched CFT can be obtained as follows:
\begin{align}
    Z^{\text{Ising}^{2*}\boldsymbol{/Z_{2}^{D}}}[\text{A},\text{A}] &= \ip{\mathcal{L}_{1}^{f}}{\psi_{1,5}}  \nonumber
    \\
    &= \me{\mathcal{L}_{1}^{f}}{\mathcal{D}_{s}}{\psi_{1,6}}  \nonumber
    \\
    &= \ip{\mathcal{L}_{1}^{f}}{\psi_{1,6}}  \nonumber
    \\
    &= Z^{\text{Ising}^{2}\boldsymbol{/Z_{2}^{D}}}[\text{A},\text{A}],
\label{eq:tgw}
\end{align}
where we have used the fact that $\bra{\mathcal{L}_{1}^{f}} \mathcal{D}_{s} = \bra{\mathcal{L}_{1}^{f}}$ in the third equality. We see that the untwisted partition function are the same as the $\text{Ising}^{2}\boldsymbol{/Z_{2}^{D}}$ CFT, which describes the transition between the trivial and the ``SSB+Kitaev" phases. 

Now we would like to show that the twisted sectors of the $\text{Ising}^{2*}\boldsymbol{/Z_{2}^{D}}$ theory is different from the $\text{Ising}^{2}\boldsymbol{/Z_{2}^{D}}$ CFT. We begin with the twisted partition function of the $\text{Ising}^{2}\boldsymbol{/Z_{2}^{D}}$ CFT with both spatial and temporal twists:
\begin{widetext}
\begin{align}
    Z^{\text{Ising}^{2}\boldsymbol{/Z_{2}^{D}}}[(\text{A},\text{P});(1,0)] &= \me{m_{1} \times \mathcal{L}_{1}^{f}}{W_{m_{1}m_{2}}}{\psi_{16}}  \nonumber
    \\
    &= Z_{m_{1}} + Z_{f_{1}e_{2}} - Z_{f_{2}} - Z_{e_{1}m_{2}}  \nonumber
    \\
    &= (|\chi_{1}|^{2} + |\chi_{\psi}|^{2})|\chi_{\sigma}|^{2} + (\chi_{1} \overline{\chi}_{\psi} + \chi_{\psi} \overline{\chi}_{1})|\chi_{\sigma}|^{2} 
    \nonumber
    \\
    &\qquad  - (|\chi_{1}|^{2} + |\chi_{\psi}|^{2})(\chi_{1} \overline{\chi}_{\psi} + \chi_{\psi} \overline{\chi}_{1}) - |\chi_{\sigma}|^{4} .
\label{eq:i2z2dcft}
\end{align}
\end{widetext}
While, in the symmetry enriched CFT $\text{Ising}^{2*}\boldsymbol{/Z_{2}^{D}}$, the partition function with the same twists is given by
\begin{widetext}
\begin{align}
    Z^{\text{Ising}^{2*}\boldsymbol{/Z_{2}^{D}}}[(\text{A},\text{P});(1,0)] &= \me{m_{1} \times \mathcal{L}_{1}^{f}}{W_{m_{1}m_{2}}}{\psi_{15}}  \nonumber
    \\
    &= \me{m_{1} \times \mathcal{L}_{1}^{f}}{W_{m_{1}m_{2}}\mathcal{D}_{s}}{\psi_{16}}  \nonumber
    \\
    &= \me{m_{1}e_{2} \times \mathcal{L}_{1}^{f}}{W_{f_{1}f_{2}}}{\psi_{16}}  \nonumber
    \\
    &= - Z_{m_{2}} - Z_{e_{1}f_{2}} + Z_{f_{1}} + Z_{m_{1}e_{2}}  \nonumber
    \\
\begin{split}
    &= - ( (|\chi_{1}|^{2} + |\chi_{\psi}|^{2})|\chi_{\sigma}|^{2} + (\chi_{1} \overline{\chi}_{\psi} + \chi_{\psi} \overline{\chi}_{1})|\chi_{\sigma}|^{2} 
    \\
    &\qquad - (|\chi_{1}|^{2} + |\chi_{\psi}|^{2})(\chi_{1} \overline{\chi}_{\psi} + \chi_{\psi} \overline{\chi}_{1}) - |\chi_{\sigma}|^{4} ) 
    \nonumber
\end{split}
    \\
    &= - Z^{\text{Ising}^{2}\boldsymbol{/Z_{2}^{D}}}[(\text{A},\text{P});(1,0)],
\label{eq:sei2z2dcft}
\end{align}
\end{widetext}
where, in the third equality, we push the twist defect $\mathcal{D}_{s}$ into the left boundary, which implements the corresponding anyon permutation for the spatial and temporal twists. We see that the twist partition functions Eq.~\eqref{eq:sei2z2dcft} and Eq.~\eqref{eq:i2z2dcft} differ by a minus sign, which comes from the SPT invariant of the Gu-Wen state. Physically, it means that the non-trivial $\zz$ flux carries an additional fermion. Similarly, one can show that the fermion parity flux is decorated by a non-trivial $\zz$ charge. The twisted partition function takes the following form:
\begin{widetext}
\begin{equation}
    Z^{\text{Ising}^{2*}\boldsymbol{/Z_{2}^{D}}}[\rho;(g,h)] = Z^{\text{Ising}\boldsymbol{/Z_{2}^{D}}}[\rho;(g,h)] Z^{\text{GW}}[\rho;(g,h)],
\end{equation}
\end{widetext}
where $Z^{GW}[\rho;(g,h)]$ is the partition function of the Gu-Wen fermionic SPT phase. This gives an example of the first type of the fermionic SEQCPs.  

\subsection{$\text{GW} \leftrightarrow \text{SSB} = \text{Ising} \times \mathcal{F}_{\text{GW}}$}
Here we discuss another example of the symmetry enriched quantum critical points. We consider the transition between the Gu-Wen and the SSB phases. In the sandwich picture, the physical boundary is tuned to be sitting at the transition between the gapped boundaries $\mathcal{L}_{4}$ and $\mathcal{L}_{5}$. We denote the corresponding boundary state as $\ket{\psi_{45}}$. The untwisted partition function is given by the overlap 
\begin{equation}
    Z^{\text{Ising}\times \mathcal{F}_{\text{GW}}}[\text{A},\text{A}] = \ip{\mathcal{L}_{1}^{f}}{\psi_{45}},
\label{eqn:zgussb}
\end{equation}
where $\mathcal{F}_{\text{GW}}$ denotes the theory of the Gu-Wen SPT phase. As we will show below, this symmetry enriched critical theory carries the topological responses coming from the Gu-Wen phase. We refer to this symmetry enriched critical theory as $\text{Ising} \times \mathcal{F}_{\text{GW}}$ with the $\zz$ symmetry in the Ising and $\mathcal{F}_{\text{GW}}$ being identified.  

To proceed, we consider the duality implemented by the defect $\mathcal{D}_{s} \circ \Tilde{\mathcal{D}}$ where $\Tilde{\mathcal{D}} = \mathcal{D}_{\sigma_{1}} \circ \mathcal{D}_{s} \circ \mathcal{D}_{\sigma_{1}\sigma_{2}}$. The defect $\Tilde{\mathcal{D}}$ acts on the gapped boundaries as 
\begin{equation}
    \Tilde{\mathcal{D}} \ket{\mathcal{L}_{1}} = \ket{\mathcal{L}_{4}}, \ \Tilde{\mathcal{D}} \ket{\mathcal{L}_{2}} = \ket{\mathcal{L}_{6}}.
\end{equation}
We can now pull the defect $\mathcal{D}_{s} \circ \Tilde{\mathcal{D}}$ from the right physical boundary and write the partition function Eq.~\eqref{eqn:zgussb} as
\begin{align}
    Z^{\text{Ising}\times \mathcal{F}_{\text{GW}}}[\text{A},\text{A}] &= \ip{\mathcal{L}_{1}^{f}}{\psi_{45}}
    \nonumber
    \\
     &= \me{\mathcal{L}_{1}^{f}}{\mathcal{D}_{s} \circ \Tilde{\mathcal{D}}}{\psi_{12}}.
\end{align}
Using the fact that the symmetry boundary $\mathcal{L}_{1}^{f}$ can absorb the gauged SPT defect $\mathcal{D}_{s}$ and action of the defect $\Tilde{\mathcal{D}}$:
\begin{equation}
    \bra{\mathcal{L}_{1}^{f}} \Tilde{\mathcal{D}} = \bra{\mathcal{L}_{5}^{f}},
\end{equation}
the partition function becomes
\begin{equation}
    Z^{\text{Ising}\times \mathcal{F}_{\text{GW}}}[\text{A},\text{A}] = \ip{\mathcal{L}_{5}^{f}}{\psi_{12}}.
\label{eqn:zl5phi12}
\end{equation}

Now we need to know the boundary state $\ket{\psi_{12}}$. We see from Table.~\ref{Table:z2z2bdy} that, on the bosonic side, this is the critical point between the partially $\zz$ ordered phase and a completely $\zz \times \zz$ symmetry breaking phase, where the remaining $\zz$ symmetry is broken spontaneously during the transition. Therefore, if we view the system as two layers of which each layer has $\zz$ symmetry, we expect that the transition only happens in the first layer and it is described by the Ising CFT. The second layer remains in the $\zz$ SSB phase during the transition. 

Upon making this assumption, we can write the boundary state $\ket{\psi_{12}}$ as
\begin{equation}
    \ket{\psi_{12}} = \sum_{\alpha,\beta \in \text{TC}} Z_{\alpha} u_{\beta} \ket{\alpha \beta},
\label{eqn:psi12}
\end{equation}
where TC is the set of anyons in the toric code, $Z_{\alpha}$ is the Ising partition function in the anyon sector $\alpha \in $ TC for the first layer, and $u_{\beta} = (1,1,0,0)$ describes the condensed anyons ($1$ and $e$) on the right physics boundary for the second layer. By taking the overlap $\ip{\mathcal{L}_{1}}{\psi_{12}}$, we recover the partition function of the Ising CFT stacking with a $\zz$ SSB gapped sector:
\begin{align}
     Z_{\mathcal{B}} = \ip{\mathcal{L}_{1}}{\psi_{12}} &= Z_{1}u_{1} + Z_{e}u_{1} + Z_{1}u_{e} + Z_{e}u_{e} \nonumber
     \\
     &=2 (|\chi_{1}|^{2} + |\chi_{\psi}|^{2} + |\chi_{\sigma}|^{2}),
\end{align}
where we have used $u_{1}=u_{e}=1$, and Eq.~\eqref{eq:Ising_TC} in the second equality, which is also called a gapless SSB phase~\cite{Bhardwaj2024hasse}.

Using the boundary state Eq.~\eqref{eqn:psi12}, we obtain the untwisted partition function for the Gu-Wen to SSB transition by calulating the overlap Eq.~\eqref{eqn:zl5phi12}:
\begin{align}
        Z^{\text{Ising}\times \mathcal{F}_{\text{GW}}}[\text{A},\text{A}] &= \ip{\mathcal{L}_{5}^{f}}{\psi_{12}} \nonumber
        \\
        &= Z_{1}u_{1} + Z_{e}u_{1}+ Z_{1}u_{f} + Z_{e}u_{f} \nonumber
        \\
        &= |\chi_{1}|^{2} + |\chi_{\psi}|^{2} + |\chi_{\sigma}|^{2},
\label{eqn:zpsi12nsns}
\end{align}
where we have used $u_{1}=1$, $u_{f}=0$, and Eq.~\eqref{eq:Ising_TC} in the second equality. Since the $\zz \times \zz^{F}$ symmetry is broken spontaneously down to $\zz^{F}$, we expect that the transition is described by an Ising CFT, which is consistent with Eq.~\eqref{eqn:zpsi12nsns}. The transition from the Gu-Wen phase to the SSB phase is in the same universality class as the critical point between the trivial and the SSB phases, of which the partition function is given by the overlap:
\begin{align}
        Z^{\text{Ising}\times \mathcal{F}_{\text{0}}}[\text{A},\text{A}] &= \ip{\mathcal{L}_{1}^{f}}{\psi_{46}} \nonumber
        \\
        &= \me{\mathcal{L}_{1}^{f}}{\Tilde{\mathcal{D}}}{\psi_{12}} \nonumber
        \\
        &= \ip{\mathcal{L}_{5}^{f}}{\psi_{12}} \nonumber
        \\
        &= |\chi_{1}|^{2} + |\chi_{\psi}|^{2} + |\chi_{\sigma}|^{2}.
\label{eqn:zgwtotrivial}
\end{align}
We note that Eq.~\eqref{eqn:zgwtotrivial} is a Ising CFT stacking with a trivial fermionic theory $\mathcal{F}_{0}$. From Eq.~\eqref{eqn:zpsi12nsns} and Eq.~\eqref{eqn:zgwtotrivial}, we confirm that the untwisted partition functions are indeed the same for both transitions. However, we will show that there are non-trivial topological responses in the twisted sectors. 

We begin with the twisted partition function of the Trivial-to-SSB transition. We consider the twist by inserting a $\zz$ defect along the time cycle and measure the fermion parity by inserting the fermion parity defect along the spatial cycle. This twisted partition function is given by the overlap:
\begin{widetext}
\begin{align}
    Z^{\text{Ising}\times \mathcal{F}_{\text{0}}}[(\text{A},\text{P});(1,0)] &= \me{m_{1} \times \mathcal{L}_{1}^{f}}{W_{m_{1}m_{2}}}{\psi_{46}}  \nonumber
    \\
    &= \me{m_{1} \times \mathcal{L}_{1}^{f}}{W_{m_{1}m_{2}} \circ \Tilde{\mathcal{D}}}{\psi_{12}}  \nonumber
    \\
    &= \me{m_{1} \times \mathcal{L}_{5}^{f}}{W_{e_{2}}}{\psi_{12}} \nonumber
    \\
    &= Z_{m}u_{1} + Z_{f}u_{1} - Z_{m}u_{f} - Z_{f}u_{f}  \nonumber
    \\
    &=  Z_{m} + Z_{f} \nonumber
    \\
    &= |\chi_{\sigma}|^{2} + \chi_{1} \overline{\chi}_{\psi} + \chi_{\psi} \overline{\chi}_{1},
\label{eq:i2dcft}
\end{align}
\end{widetext}
where, in the fifth equality, we have used $u_{1} = u_{e} = 1$ and $u_{f} = u_{m} = 0$. Eq.~\eqref{eq:i2dcft} is the same as the twisted partition function $Z_{1,0}$ in Eq.~\eqref{eqn:zising10} with only an insertion of the $\zz$ defect along the time cycle in the Ising CFT. This is expected since the Ising CFT is essentially decoupled from the trivially gapped fermionic sector $\mathcal{F}_{0}$.

Now we consider the same twist in the $\text{Ising} \times \mathcal{F}_{\text{GW}}$ theory, which is given by 
\begin{widetext}
\begin{align}
    Z^{\text{Ising}\times \mathcal{F}_{\text{GW}}}[(\text{A},\text{P});(1,0)] &= \me{m_{1} \times \mathcal{L}_{1}^{f}}{W_{m_{1}m_{2}}}{\psi_{45}} \nonumber 
    \\
    &= \me{m_{1} \times \mathcal{L}_{1}^{f}}{W_{m_{1}m_{2}} \circ \mathcal{D}_{s} \circ \Tilde{\mathcal{D}}}{\psi_{12}}  \nonumber
    \\
    &= \me{m_{1}e_{2} \times \mathcal{L}_{5}^{f}}{W_{e_{1}e_{2}}}{\psi_{12}} \nonumber
    \\
    &= Z_{m}u_{m} + Z_{f}u_{m} - Z_{m}u_{e} - Z_{f}u_{e} \nonumber 
    \\
    &= - (Z_{m} + Z_{f}) \nonumber
    \\
    &= -( |\chi_{\sigma}|^{2} + \chi_{1} \overline{\chi}_{\psi} + \chi_{\psi} \overline{\chi}_{1}) \nonumber
    \\
    &= - Z^{\text{Ising}\times \mathcal{F}_{\text{0}}}[(\text{A},\text{P});(1,0)].
\label{eq:i2gwdcft}
\end{align}
\end{widetext}
The minus sign in Eq.~\eqref{eq:i2gwdcft} is the topological invariant coming from the Gu-Wen phase. It means that the $\zz$ flux in the Ising CFT is decorated by an addition fermion from the gapped sector. Similarly, one can show that the fermion parity flux in the gapped sector is decorated by a non-trivial $\zz$ charge in the Ising CFT. This is also an example of the first type of the fermionic SEQCPs and can be understood as the stacking of the GW-Wen ferminic SPT state.

\subsection{($\text{Kitaev} + \text{GW}) \leftrightarrow \text{SSB} = (\text{Ising} \times \text{Majorana})^{*} $}

Here we discuss a new type of the symmetry enriched quantum critical point. We consider the transition between the ``\text{Kitaev} + \text{GW}" phase and the $\zz$ SSB phases. This corresponds to the transition between the $\mathcal{L}_{3}$ and $\mathcal{L}_{4}$ boundaries on the physical boundary and the symmetry boundary is still $\mathcal{L}_{1}^{f}$. We denote the corresponding critical boundary state as $\ket{\psi_{3,4}}$ and the partition function of the SPT transition is $Z[{\text{A},\text{A}}] = \ip{\mathcal{L}_{1}^{f}}{\psi_{3,4}}$. 

We would like to relate the critical boundary state $\ket{\psi_{3,4}}$ to $\ket{\psi_{1,6}}$, which we know is described by the $\text{Ising}^{2}$ CFT. To perform the duality, we consider the defect $\mathcal{D}_{\sigma_{2}} \circ \mathcal{D}_{s}$ such that the action is given by $\mathcal{D}_{\sigma_{2}} \circ \mathcal{D}_{s} \ket{\psi_{1,6}} = \ket{\psi_{3,4}}$. 
The untwisted partition function for the transition is given by
\begin{widetext}
\begin{align}
    Z[\text{A},\text{A}] &= \ip{\mathcal{L}_{1}^{f}}{\psi_{3,4}}  \nonumber
    \\
    &= \me{\mathcal{L}_{1}^{f}}{\mathcal{D}_{\sigma_{2}} \circ \mathcal{D}_{s}}{\psi_{1,6}} \nonumber
    \\
    &= \ip{\mathcal{L}_{7}^{f}}{\psi_{1,6}} \nonumber
    \\
    &= (Z_{1} + Z_{m})(Z_{1} + Z_{f}) \nonumber
    \\
    &=  (|\chi_{1}|^{2} + |\chi_{\sigma}|^{2} + |\chi_{\psi}|^{2}) (|\chi_{1}|^{2} + |\chi_{\psi}|^{2} + \chi_{1} \overline{\chi}_{\psi} + \chi_{\psi} \overline{\chi}_{1}),
\label{eq:tgwz4}
\end{align}
\end{widetext}
which is the same as the Ising $\times$ Majorana CFT. 

The twisted partition functions however differ from the Ising $\times$ Majorana CFT as we now show. We consider the twisted partition function with a $\zz$ twist along the spatial direction and the fermion parity twist along the time direction, which is given by
\begin{widetext}
\begin{align}
    Z[(\text{A},\text{P});(1,0)] &= \me{m_{1} \times \mathcal{L}_{1}^{f}}{W_{m_{1}m_{2}}}{\psi_{3,4}} \nonumber 
    \\
    &= \me{m_{1} \times \mathcal{L}_{1}^{f}}{W_{m_{1}m_{2}} \circ \mathcal{D}_{2} \circ \mathcal{D}_{s} }{\psi_{1,6}}  \nonumber
    \\
    &= \me{m_{1}e_{2} \times \mathcal{L}_{7}^{f}}{W_{m_{1}}}{\psi_{1,6}} \nonumber
    \\
    &= (Z_{e}+Z_{f})(Z_{m}-Z_{e}) \nonumber 
    \\
    &= Z^{\text{Majorana}}[(\text{P},\text{P})] Z^{\text{Ising}}[(1,0)]  \nonumber
    \\
    &= Z[(\text{P},\text{P});(1,0)].
\label{eq:i2gwdcft2}
\end{align}
\end{widetext}
We find that the $\zz$ twist along the spatial direction induces a change of the spin structure along the same direction. This happens since the $\zz$ defect line is attached with an additional $\zz^{F}$ fermion parity line along the same direction, which causes the change of the spin structure. This is an example of the second type of the fermionic SEQCPs, which can be understood as performing the twist gauging $\textbf{S}\textbf{T}$ for the $\text{Ising}^{2*}\boldsymbol{/Z_{2}^{D}}$ theory where $\textbf{S}$ here is the gauging of the $\zz$ symmetry in $\zz \times \zz^{F}$. 

\section{Fermionic chains with $\mathbb{Z}_{4}^{F}$ symmetry}
\label{sec:z4f}

We discuss the quantum criticality in the fermionic chains with $\mathbb{Z}_{4}^{F}$ symmetry in this section. In the sandwich construction, we choose the bulk to be the $\z_{4}$ gauge theory (also called the $\z_{4}$ toric code topological order). 

We briefly review the basic properties of the $\z_{4}$ toric code topological order below. The anyons satisfy a $\z_{4} \times \z_{4}$ fusion group, generated by the charge $e$ and the flux $m$. There is a non-trivial mutual braiding statistics between $e$ and $m$
\begin{equation}
    \frac{S_{e,m}}{S_{e,0}} =\frac{S_{e,m}}{S_{0,m}} = i.
\end{equation}
The self-statistics of a generic anyon $e^{p}m^{q}$ with $p,q \in \z_{4}$ is given by
\begin{equation}
    \theta(e^{p}m^{q}) = i^{pq}.
\end{equation}
The topological symmetry of the anyon theory is $\mathbb{Z}_{2} \times \mathbb{Z}_{2}$. The first $\mathbb{Z}_{2}$ is generated by
\begin{equation}
    \sigma: m \leftrightarrow e, 
\label{eq:autosigma}
\end{equation}
which corresponds to the EM exchange automorphism of the $\mathbb{Z}_{4}$ toric code. The other automorphism is given by 
\begin{equation}
    s: e \rightarrow e^{3}, m \rightarrow m^{3}
\label{eq:autos}
\end{equation}
which comes from the automorphism of the $\mathbb{Z}_{4}$ group. The automorphism $s$ can also be viewed as the charge conjugation symmetry of the $\z_4$ toric code. There are three gapped boundaries with the following Lagrangian algebra of condensed anyons
\begin{align}
    \mathcal{L}_{1} &= 1\oplus e\oplus e^{2}\oplus e^{3}, \nonumber
    \\
    \mathcal{L}_{2} &= 1\oplus e^{2}\oplus m^{2}\oplus e^{2}m^{2}, \nonumber
    \\
    \mathcal{L}_{3} &= 1\oplus m\oplus m^{2}\oplus m^{3}.
\end{align}
Choosing the symmetry boundary to be $\mathcal{L}_{1}$ corresponds to the bosonic systems with $\z_{4}$ symmetry, while choosing $\mathcal{L}_{2}$ corresponds to the bosonic systems with an anomalous $\zz \times \zz$ symmetry with the mixed anomaly in $\mathcal{H}^{2}(\mathbb{Z}_{2},\mathcal{H}^{1}(\mathbb{Z}_{2},\U))$. The second choice describes the $\zz \times \zz$ deconfined quantum critical points~\cite{Chatterjee2023holo,Huang2023symtft}.

In order to describe fermionic systems, we need to know the fermionic gapped boundaries of the $\z_{4}$ toric code. There are two super Lagrangian algebra:
\begin{align}
     \mathcal{L}_{1}^{f} &= 1\oplus em^{2} \oplus e^{2}\oplus e^{3}m^{2}, \nonumber
     \\
     \mathcal{L}_{2}^{f} &= 1\oplus e^{2}m \oplus m^{2}\oplus e^{2}m^{3}.
\end{align}
We choose $\mathcal{L}_{1}^{f}$ as the symmetry boundary, which corresponds to the fermionic systems with $\z_{4}^{F}$ symmetry. The $\z_{4}^{F}$ symmetry is generated by the $m$ line on the symmetry boundary. The fermion parity symmetry is the $\zz$ normal subgroup in $\z_{4}^{F}$, which is generated by the $m^{2}$ line:
\begin{align}
    \z_{4}: m,  \quad   \zz^{F}: m^{2}.
\label{eqn:z4fsymlines}
\end{align}
The mapping from the bulk lines to the boundary lines is given by:
\begin{align}
    \{ m, em^{3}, e^{2}m, e^{3}m^{3} \} &\rightarrow m, \nonumber
    \\
    \{ m^{2}, e, e^{2}m^{2}, e^{3} \} &\rightarrow m^{2}.
\end{align}
We note that the other choice of the symmetry boundary $\mathcal{L}_{2}^{f}$ is related by the EM exchange automorphism. 

\begin{table*}[t]\centering
	\begin{tabular}{c|c}
		Lagrangian algebra & Fermionic gapped phases
		\\
		\hline
		$\mathcal{L}_{1} = 1\oplus e\oplus e^{2}\oplus e^{3}$  & $\zz$ SSB + Kitaev
		\\ 
		\hline
		$\mathcal{L}_{2} = 1\oplus e^{2}\oplus m^{2}\oplus e^{2}m^{2}$  & $\zz$ SSB 
		\\ 
		\hline
		$\mathcal{L}_{3} = 1\oplus m\oplus m^{2}\oplus m^{3}$  & $\z_{4}^{F}$ Trivial
		\\ 
		\hline
	\end{tabular}
	\caption{Fermionic gapped phases with $\z_{4}^{F}$ symmetry. The left symmetry boundary is chosen to be the fermionic gapped boundary $\mathcal{L}^{f}_{1}$. The first column shows the physical boundary on the right of the sandwich. The second column shows the fermoinic gapped phases realized by the sandwich construction. ``$\zz$ SSB" means that the $\z_{4}^{F}$ symmetry is broken spontaneously down to $\zz^{F}$. ``Kitaev" denotes the fermionic chain is in the Kitaev phase. ``$\z_{4}^{F}$ Trivial" means the fermoinic chain is in the $\z_{4}^{F}$ symmetric trivial phase.
    }
	\label{Table:z4fgapped}
\end{table*}

\subsection{Fermionization}
\label{sec:fermionization-z4}
If we start from a bosonic system with $\z_{4}$ symmetry by choosing the symmetry boundary to be the e-condensed boundary and change the symmetry boundary to be the fermionic boundary $\mathcal{L}_{1}^{f}$, this process implements the fermionization with respecting to the $\zz$ subgroup in the $\z_{4}$ symmetry. Here we derive the fermionization formula in the topological holography. To proceed, we need the following twisted partition functions on the bosonic side:
\begin{align}
    Z_{0,0} &= Z_{1} + Z_{e} + Z_{e^{2}} + Z_{e^{3}} , \nonumber
    \\
    Z_{0,2} &= Z_{1} - Z_{e} + Z_{e^{2}} - Z_{e^{3}} , \nonumber
    \\
    Z_{2,0} &= Z_{m^{2}} + Z_{em^{2}} + Z_{e^{2}m^{2}} + Z_{e^{3}m^{2}} , \nonumber
    \\
    Z_{2,2} &= Z_{m^{2}} - Z_{em^{2}} + Z_{e^{2}m^{2}} - Z_{e^{3}m^{2}} ,
\label{eqn:z4tz2}
\end{align} 
where we have used Eq.~\eqref{eq:ztwistanyon}. In the sandwich construction, the fermionic partition function in the $(\text{A},\text{A})$ spin structure is defined as the partition function without inserting the fermion parity line, which can be written in the anyon basis:
\begin{equation}
    Z_{\mathcal{F}}[\text{A},\text{A}] = Z_{1} + Z_{em^{2}} + Z_{e^{2}} + Z_{e^{3}m^{2}}.
\label{eqn:z4fznsnsanyon}
\end{equation}
We can rewrite Eq.~\eqref{eqn:z4fznsnsanyon} by using Eq.~\eqref{eqn:z4tz2}:
\begin{equation}
   Z_{\mathcal{F}}[\text{A},\text{A}] = \frac{1}{2} (Z_{0,0} + Z_{0,2} + Z_{2,0} - Z_{2,2}).
\end{equation}
Similarly, we obtain the following fermionic partition functions in different spin structures:
\begin{align}
    Z_{\mathcal{F}}[\text{A},\text{P}] &= Z_{1} - Z_{em^{2}} + Z_{e^{2}} - Z_{e^{3}m^{2}} \nonumber
    \\
    &= \frac{1}{2} (Z_{0,0} + Z_{0,2} - Z_{2,0} + Z_{2,2}), \nonumber
    \\
    Z_{\mathcal{F}}[\text{P},\text{A}] &= Z_{e} + Z_{e^{2}m^{2}} + Z_{e^{3}} + Z_{m^{2}} \nonumber
    \\
    &= \frac{1}{2} (Z_{0,0} - Z_{0,2} + Z_{2,0} + Z_{2,2}), \nonumber
    \\
    Z_{\mathcal{F}}[\text{P},\text{P}] &= - Z_{e} + Z_{e^{2}m^{2}} - Z_{e^{3}} + Z_{m^{2}} \nonumber
    \\
    &= \frac{1}{2} (- Z_{0,0} + Z_{0,2} + Z_{2,0} + Z_{2,2}),
\label{eqn:z4tzspin}
\end{align}
which are precisely given by the usual fermionization formula Eq.~\eqref{eq:fermionization-2} with respecting to the $\zz$ subgroup \footnote{We note that, in calculating the twisted partition functions, it's possible to choose a bulk representative of the fermion parity line from the set of bulk anyon lines $\{ m^{2}, e, e^{2}m^{2}, e^{3} \}$. All these choices give consistent results once the half-braiding phase factors are taking into account.}.

If we start from a bosonic system with the anomalous $\zz \times \zz$ symmetry with the mixed anomaly by choosing the symmetry boundary to be $\mathcal{L}_{2}$ and change the symmetry boundary to be the fermionic boundary $\mathcal{L}_{1}^{f}$, the result is that it implements the other fermionization procedure with an additional stacking of the Kitaev chain. This result can be derived by following the similar procedure with the help of the following twisted partition functions:
\begin{align}
    Z_{(0,0),(0,0)} &= Z_{1} + Z_{e^{2}} + Z_{m^{2}} + Z_{e^{2}m^{2}} , \nonumber
    \\
    Z_{(0,0),(1,0)} &= Z_{1} + Z_{e^{2}} - Z_{m^{2}} - Z_{e^{2}m^{2}} , \nonumber
    \\
    Z_{(1,0),(0,0)} &= Z_{e} + Z_{e^{3}} + Z_{em^{2}} + Z_{e^{3}m^{2}} , \nonumber
    \\
    Z_{(1,0),(1,0)} &= Z_{e} + Z_{e^{3}} - Z_{em^{2}} - Z_{e^{3}m^{2}}.
\label{eqn:z4tz2z2a}
\end{align} 
For example, by using Eq.~\eqref{eqn:z4fznsnsanyon} and  Eq.~\eqref{eqn:z4tz2z2a}, the fermionic partition function in the $(\text{P},\text{P})$ spin structure is given by
\begin{widetext}
\begin{align}
        Z_{\mathcal{F}}[\text{P},\text{P}] &= -Z_{e} + Z_{e^{2}m^{2}} - Z_{e^{3}} + Z_{m^{2}} \nonumber
        \\
        &= - \frac{1}{2}(-Z_{(0,0),(0,0)} + Z_{(0,0),(1,0)} + Z_{(1,0),(0,0)} + Z_{(1,0),(1,0)}),
\label{eqn:z2z2anofznsnsanyon}
\end{align}
\end{widetext}
which differs from the fermionization formula Eq.~\eqref{eq:fermionization-2} by a minus sign, coming from the additional Arf invariant. Following similar procedure, one can check the fermionization formula for the other spin structures explicitly. We note that the anomalous $\zz \times \zz$ symmetry becomes an anomaly free $\z_{4}^{F}$ symmetry generated by the lines in Eq.~\eqref{eqn:z4fsymlines} after the fermionization.

\subsection{Gapped phases}
By choosing the physical boundary to be $\mathcal{L}_{i}$ for $i=1,2,3$, the sandwich construction realizes fermionic gapped phases with $\z_{4}^{F}$ symmetry. There are only three gapped phases: the $\z_{4}^{F}$ symmetric trivial phase, the ``$\zz$ SSB" phase such that the $\z_{4}^{F}$ symmetry is broken spontaneously down to $\zz^{F}$ symmetry, and the ``$\zz$ SSB + Kitaev" phase where the $\z_{4}^{F}$ symmetry is broken spontaneously down to $\zz^{F}$ and the system is in the Kitaev phase. 

The identification of the fermionic gapped phases can be obtained by calculating the twisted partition functions by using Eq.~\eqref{eq:ztwistanyon} and Eq.~\eqref{eq:ztwistanyon-f}. The fermionic boundary $\mathcal{L}_{3}$ corresponds to the $\z_{4}^{F}$ symmetric trivial phase since the identity line $\mathcal{W}_{1}$ is the only local operator. We check that all the twisted partition functions equal to $1$. 

The remaining boundaries $\mathcal{L}_{1}$ and $\mathcal{L}_{2}$ break the symmetry down to $\zz^{F}$ since there is a non-trivial $\zz$ order parameter given by $\mathcal{W}_{e^{2}}$. 

To see that $\mathcal{L}_{1}$ is in the Kitaev phase, we calculate the partition function in the $(\text{P},\text{P})$ spin structure by using Eq.~\eqref{eqn:z4tzspin} and we obtain
$Z_{\mathcal{F}}[\text{P},\text{P}] = -Z_{e} - Z_{e^{3}} = -2$. The absolute value ``2" indicates that there are two degenerate ground states, which is consistent with the $\zz$ SSB. The $(-1)$ sign for the $(\text{P},\text{P})$ spin structure is the signature of the Kitaev phase. For $\mathcal{L}_{2}$, there is no addition minus sign in the  $(\text{P},\text{P})$ spin structure: $Z_{\mathcal{F}}[\text{P},\text{P}] =  Z_{e} + Z_{e^{3}} = 2$, which means that it's the $\zz$ SSB phase. The identification of the fermionic gapped phases are summarized in Table.~\ref{Table:z4fgapped}.

\subsection{Quantum critical points}
\subsubsection{$\z_{4}^{F} \ \text{Trivial} \leftrightarrow (\zz \ \text{SSB} + \text{Kitaev}) = \text{Majorana}^{2} $}

\begin{figure}
	\centering
	\includegraphics[width=0.48\textwidth]{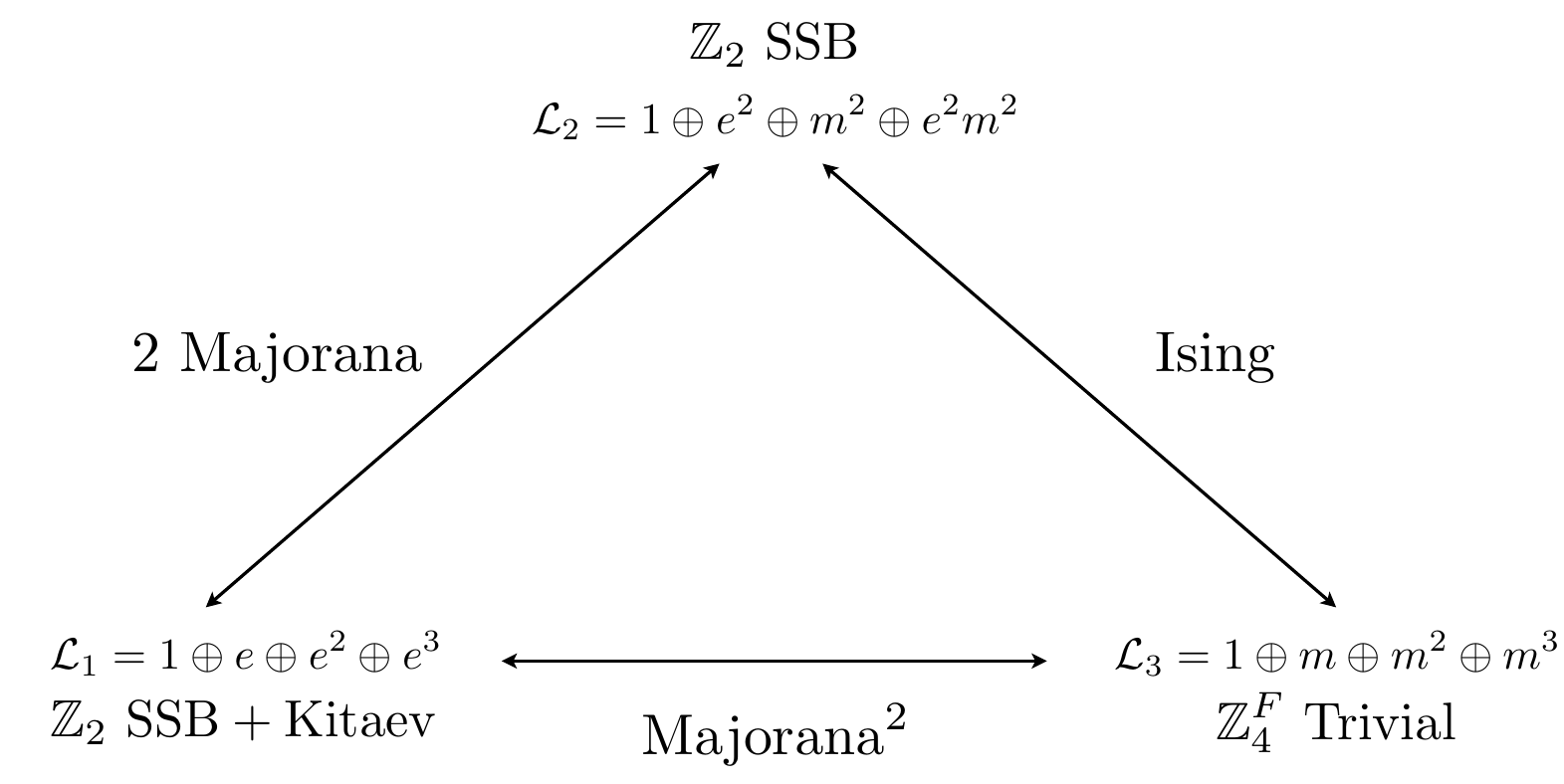}
	\caption{Summary of the phase transitions in the fermionic systems with $\z_{4}^{F}$ symmetry.}
	\label{fig:z4ftrans}
\end{figure}

Topological holography can be used to find the CFT description between any of the gapped phases identified in the previous sections. To illustrate the idea, we focus on the transition between the $\z_{4}^{F}$ symmetric trivial phase and the ``$\zz$ SSB + Kitaev" phase. The input we are using is the $U(1)_{4}$ CFT, which can be the DQCP transition for the anomalous $\zz \times \zz$ symmetry with the mixed anomaly. 

Recall that, in the sandwich picture of the DQCP, we choose the symmetry boundary to be $\mathcal{L}_{2}$ and tune the physical boundary to be sitting at the transition between $\mathcal{L}_{1}$ and $\mathcal{L}_{3}$, which we assume to be described by the $U(1)_{4}$ CFT. Now we perform the fermionization by changing the symmetry boundary to be $\mathcal{L}_{1}^{f}$. From the discussion of Sec.~\ref{sec:fermionization-z4}, we see that it corresponds to fermionizing with respect to the first $\zz$ symmetry. By applying Eq.~\eqref{eq:fermionization-2}, we obtain the fermionic partition function in the $(\text{A},\text{A})$ spin structure:
\begin{align}
    Z_{\mathcal{F}}[\text{A},\text{A}] &= (|K_{0}|^{2} + |K_{2}|^{2} + K_{0}\bar{K}_{2} + K_{2}\bar{K}_{0} ), \nonumber
    \\
    &= (|\chi_{1}|^{2} + |\chi_{\psi}|^{2} + \chi_{1}\bar{\chi}_{\psi} + \chi_{\psi}\bar{\chi}_{1} )^{2},
\label{eqn:zMaj2}
\end{align}
where $K_{n}$ with $n=-1,0,1,2$ are the characters of the $U(1)_{4}$ CFT, and $\chi_{a}$ with $a=\{1,\sigma,\psi \}$ are the characters in the Ising CFT. We recognize that Eq.~\eqref{eqn:zMaj2} is the partition function for two copies of the Majorana CFT, which is equivalent to a Dirac theory. We note that Ref.~\cite{Gaiotto2021} also discuss fermionizing the compact boson theory to obtain a Dirac theory and the appearance of the $\z_{4}^{F}$ symmetry by using pure CFT analysis. 

\subsubsection{$\z_{4}^{F} \ \text{Trivial} \leftrightarrow \zz \ \text{SSB} = \text{Ising} \times \mathcal{F}_{0} $}
The CFT description for other critical points can be obtain by using the ``club sandwich" construction~\cite{Bhardwaj2023club}. To study the transition between the $\z_{4}^{F}$ trivial and the $\zz$ SSB phases, we consider the following club sandwich. The bulk of the club sandwich consists of a $\z_{4}$ toric code topological order on the left and a $\zz$ toric code topological order on the right, separating by a gapped domain wall on which the $m^{2}$ particle in the $\z_{4}$ toric code condenses. The particle $e^{2}$ and $m$ (and their fusion products with the condensate $m^{2}$) becomes the $e'$ and $m'$ particles in the $\zz$ toric code. We will denote the domain wall by $\mathcal{I}_{m^{2}}$. If the physical boundary is the $e'$-condensed boundary $\mathcal{A}_{e'}$ of the $\zz$ toric code, the composition between the domain wall $\mathcal{I}_{m^{2}}$ and the boundary $\mathcal{A}_{e'}$ gives the $\mathcal{L}_{2}$ boundary in the $\z_{4}$ toric code. Similarly, the composition between the domain wall $\mathcal{I}_{m^{2}}$ and the $m$-condensed boundary $\mathcal{A}_{m'}$ gives the $\mathcal{L}_{3}$ boundary. We then tune the physical boundary to be sitting at the transition between the $e'$ and $m'$ condensed boundaries, which is described by the Ising CFT. The fermionization is again implemented by changing the symmetry boundary to be $\mathcal{L}_{1}^{f}$. The partition function is given by the overlap
\begin{align}
     Z^{\text{Ising}\times \mathcal{F}_{\text{0}}}[\text{A},\text{A}] &= \mel{\mathcal{L}_{1}^{f}}{\mathcal{I}_{m^{2}}}{\Psi_{\text{Ising}}} \nonumber
     \\
     &= Z_{1} + Z_{e'}, \nonumber
     \\
     &= |\chi_{1}|^{2} + |\chi_{\psi}|^{2} + |\chi_{\sigma}|^{2},
\end{align}
which is the Ising CFT stacking with a trivial fermionic theory $\mathcal{F}_{0}$. 

\subsubsection{$\zz \ \text{SSB} \leftrightarrow (\zz \ \text{SSB} + \text{Kitaev}) = 2 \ \text{Majorana} $}
For the transition between the $\zz$ SSB and the ``$\zz$ SSB + Kitaev" phases, we consider the similar club sandwich consists of the $\z_{4}$ toric code and the $\zz$ toric code, separated by a gapped domain wall on which the $e^{2}$ particle in the $\z_{4}$ toric code condenses. We will denote the domain wall by $\mathcal{I}_{e^{2}}$. The particle $e$ and $e^{3}$ becomes the $e'$ particle and $m^{2}$ becomes the $m'$ particles in the $\zz$ toric code. We also tune the physical boundary to be sitting at the transition between the $e'$ and $m'$ condensed boundaries, which is described by the Ising CFT and change the symmetry boundary to be $\mathcal{L}_{1}^{f}$ to perform the fermionization. The partition function is given by
\begin{widetext}
\begin{align}
     Z^{2 \text{Majorana}}[\text{A},\text{A}] &= \mel{\mathcal{L}_{1}^{f}}{\mathcal{I}_{e^{2}}}{\Psi_{\text{Ising}}} \nonumber
     \\
     &= 2(Z_{1} + Z_{f'}) , \nonumber
     \\
     &=2( |\chi_{1}|^{2} + |\chi_{\psi}|^{2} + \chi_{1} \overline{\chi}_{\psi} + \chi_{\psi} \overline{\chi}_{1}) \nonumber
     \\
     &= 2 Z^{\text{Majorana}}[\text{A},\text{A}],
\end{align}
\end{widetext}
where the prefactor ``$2$" in the second equality comes from the summation over the condensed particles $\{ 1, m^{4}, em^{2}, e^{3}m^{2} \}$ on the symmetry boundary $\mathcal{L}_{1}^{f}$ which can pass through the gapped domain wall $\mathcal{I}_{e^{2}}$. Physically, it signatures the 2-fold ground state degeneracy since the $\z_{4}^{F}$ symmetry has already broken down to $\zz^{F}$. The transition is given by the free Majorana CFT in the two superselection sectors, which is also called a gapless SSB phase~\cite{Bhardwaj2024hasse}. This theory has the same primaries as the free Majorana CFT but every level in the spectrum is at least two-fold degenerate. This concludes the discussion of the phase transitions in the fermionic systems with $\z_{4}^{F}$ symmetry and the results are summarized in Fig.~\ref{fig:z4ftrans}.

\subsection{The fermionic intrinsically gapless SPT phase with $\z_{4}^{F}$ symmetry}
Here we discuss an example of the intrinsically gapless fermionic SPT with $\z_{4}^{F}$ symmetry. We first briefly review the basic idea of the bosonic igSPT phases.

An bosonic igSPT phase is a gapless phase that exhibiting an emergent IR anomaly in a bosonic lattice model in $D-1$ spatial dimensions with an anomaly-free UV symmetry~\cite{Thorngren2021igspt,Potter2022}. In general we have the following symmetry structure. Let $\Gamma$ be the microscopic on-site symmetry, which is anomaly free. Below an energy scale $\Delta$, there is an emergent IR symmetry $G$ that sits in the short exact sequence:
\begin{equation}
    1\rightarrow H\rightarrow \Gamma \rightarrow G\rightarrow 1,
\end{equation}
where $H$ is a normal subgroup of $\Gamma$ which acts trivially on the IR degrees of freedom, and $G = \Gamma / H$. The emergent IR symmetry $G$ has an anomaly in $\mathcal{H}^{D+1}(G,\U)$, where $D$ is the spacetime dimensions. This anomaly becomes trivial when it is pulled back into $\mathcal{H}^{D+1}(\Gamma,\U)$, so that the entire system can be realized in the lattice models $D-1$ spatial dimensions with non-anomalous $\Gamma$ symmetry.

An simple example is given by the $(1+1)d$ igSPT phases with $\Gamma = \mathbb{Z}_{4}$, $H = \mathbb{Z}_{2}$, and $G = \mathbb{Z}_{2}$. The non-trivial igSPT phase realized in the $\Gamma = \mathbb{Z}_{4}$ bosonic lattice model shares the same (emergent) anomaly as the gapless edge modes of the $\mathbb{Z}_{2}$ Levin-Gu SPT state. 

The sandwich picture of the $\z_{4}$ igSPT phase is given by the club sandwich construction as discussed in Ref.~\cite{Huang2023symtft,Bhardwaj2023club}. We have a $\mathbb{Z}_{4}$ toric code topological order on the left and a double semion (DS) order on the right. The two topological orders are separated by a gapped domain wall on which the $e^{2}m^{2}$ particle in the $\mathbb{Z}_{4}$ toric code condenses. We will denote the domain wall by $\mathcal{I}_{e^2m^2}$. After the $e^{2}m^{2}$ condensation, the deconfined anyons in the $\mathbb{Z}_{4}$ toric code are $\{ [1], [em], [em^{3}], [m^{2}] \}$, which can be identified with the anyons in the double semion order $\{ 1, s, \bar{s}, b \}$, respectively. For the bosonic $\z_{4}$ igSPT phase, the symmetry boundary is chosen to be the charge condensed boundary $\mathcal{L}_{1}$.

Since $H = \mathbb{Z}_{2}$ is non-anomalous, we can perform the fermionization with respecting the $H = \mathbb{Z}_{2}$ subgroup. This corresponds to change the symmetry boundary to be the fermionic gapped boundary $\mathcal{L}_{1}^{f}$ in the club sandwich construction and the global symmetry becomes $\z_{4}^{F}$. 

On the physical boundary of the club sandwich, we have an edge theory of the $\mathbb{Z}_{2}$ Levin-Gu SPT state, which can live at the edge of the DS order. Upon dimensional reduction, the sandwich construction realizes the igSPT state where the low-energy theory realizes the anomalous edge theory of the $\mathbb{Z}_{2}$ Levin-Gu SPT state in a system with a $\mathbb{Z}_{4}^{F}$ global symmetry. 

As a concrete example, we can choose the physical boundary to be the $\U_{2}$ CFT~\cite{Ji2019}. The partition functions in the DS sectors are given by:
\begin{align}
    Z_{1} &= |\chi_{0}|^{2}, \nonumber
    \\
    Z_{s} &= \chi_{1}\overline{\chi}_{0}, \nonumber
    \\
    Z_{\bar{s}} &= \chi_{0}\overline{\chi}_{1}, \nonumber
    \\
    Z_{b} &= |\chi_{1}|^{2}.
\label{eq:DS_u12}
\end{align}
The gapped domain wall $\mathcal{I}_{e^{2}m^{2}}$ in the club sandwich construction can be written as
\begin{equation}
    \mathcal{I}_{e^{2}m^{2}} = \sum_{\alpha \in \mathbb{Z}_{4} \text{TC},\mu \in \text{DS}} W_{\alpha ,\mu} \ket{\alpha} \bra{\mu},
\end{equation}
where $W_{\alpha ,\mu}=1$ only for transparent anyons. The untwisted torus partition function can then be obtained as the overlap with the insertion of the gapped domain wall $\mathcal{I}_{e^{2}m^{2}}$:
\begin{align}
     Z_{00} &= \mel{\mathcal{L}_{1}^{f}}{\mathcal{I}_{e^{2}m^{2}}}{\Psi} \nonumber
     \\
     &=  \sum_{\alpha \in \mathbb{Z}_{4} \text{TC},\mu \in \text{DS}} w_{\alpha} W_{\alpha ,\mu} Z_{\mu} \nonumber
     \\
     &= Z_{1} + Z_{b},
     \\
     &= |\chi_{0}|^{2}+|\chi_{1}|^{2}.
\end{align}
where in the second equality $w_{\alpha} = (1,0,1,0)$ indexed as $(w_{1},w_{em^{2}},w_{e^{2}},w_{e^{3}m^{2}})$. We find that the untwisted partition function is the same as the usual $\U_{2}$ CFT.

The non-trivial characterization is in the twisted partition functions since we can twist by the element in $\zz^{F}$. The analysis is parallel to the bosonic $\z_{4}$ igSPT phase by replacing the $\zz$ normal subgroup to $\zz^{F}$~\cite{Bhardwaj2023club}. Here we only give an example to illustrate the main idea and other twisted partition functions can be derived similarly. Consider the partition functions with a $g \in G = \zz$ twist in the space direction and a $h \in H = \zz^{F}$ twist in the time direction:
\begin{equation}
    Z_{g,h} = \Tr_{\mathcal{H}_{g}} h e^{-\beta \mathcal{H}}. 
\end{equation}
Recall that the $g$ twist corresponds to an $m$ line on the left boundary that generates the $\Gamma = \z_{4}^{F}$ symmetry. The $h$ twist corresponds to the $m^{2}$ line that generates the $\zz^{F}$ normal subgroup in $\Gamma = \z_{4}^{F}$.
The twisted partition function can be calculated in the club sandwich as
\begin{align}
    Z_{g,h} &= \sum_{\alpha \in m \times \mathcal{L}_{e} ,\mu \in \text{DS}}  \frac{S_{\alpha,m^{2}}}{S_{0,\alpha}}  W_{\alpha ,\mu} Z_{\mu} \nonumber
    \\
    &= - (Z_{s}+ Z_{\bar{s}}) \nonumber
    \\
    &= - Z_{g,0},
\label{eq:igSPT12twist}
\end{align}
where $m \times \mathcal{L}_{e} = (m,em^{3},e^{2}m,e^{3}m^{3})$. The minus sign in Eq.~\eqref{eq:igSPT12twist} means that there is an additional fermion decorated on the $g$ defect line, and the twisted partition function $Z_{g,h}$ precisely measures the fermion parity in the presence of $g$ defect. 

\section{Fermionic chains with super fusion category $\text{SRep}(\mathcal{H}_{S_{3}})$ symmetry}
\label{sec:fnoninv}

In this section we focus on fermionic systems with non-invertible symmetries. In particular, we consider 1D fermionic systems with the super fusion category $\text{SRep}(\mathcal{H}_{S_{3}})$ symmetry, where $\mathcal{H}_{S_{3}}$ is a Hopf superalgebra~\cite{Inamura2023}. This symmetry emerges out of the fermionization with respecting to the $\zz$ symmetry in the $\text{Rep}(S_{3})$ symmetry. We give summary of the sandwich picture of the bosonic systems with $S_{3}$ and $\text{Rep}(S_{3})$ symmetry below.

We begin with a review for the bosonic systems with $S_3=\{r,s|r^3=s^2=1, srs=r^{-1}\}$ symmetry. The bulk of the sandwich is the $S_3$ gauge theory with 8 anyon types.  Following the convention in Ref.~\cite{Cui:2014sfa}, we label them as $A, B, \dots, H$. $A,B$ and $C$ are the three gauge charges: $A$ corresponds to the identity representation, $B$ is the one-dimensional representation and $C$ is the two-dimensional representation. $D$ and $E$ are the two fluxes with conjugacy class $[s]=\{s,sr,sr^2\}$. $F,G,H$ are the three fluxes with conjugacy class $[r]=\{r,r^2\}$. The $S$-matrix of the $S_{3}$ gauge theory is 
\begin{equation}
    S = \frac{1}{6} \begin{pmatrix}
	1 & 1 & 2 & 3 & 3 & 2 & 2 & 2 \\
    1 & 1 & 2 & -3 & -3 & 2 & 2 & 2 \\
    2 & 2 & 4 & 0 & 0 & -2 & -2 & -2 \\
    3 & -3 & 0 & 3 & -3 & 0 & 0 & 0 \\
    3 & -3 & 0 & -3 & 3 & 0 & 0 & 0 \\
    2 & 2 & -2 & 0 & 0 & 4 & -2 & -2 \\
    2 & 2 & -2 & 0 & 0 & -2 & -2 & 4 \\
    2 & 2 & -2 & 0 & 0 & -2 & 4 & -2 \\
	\end{pmatrix}.
\label{eq:smatrixs3}
\end{equation}
The topological spins are 
\begin{equation}
    \{ 0,0,0,0,\frac{1}{2},0,\frac{1}{3},\frac{2}{3} \},
\end{equation}
for anyons $A, B, \dots, H$. There is a non-trivial automorphism given by exchanging $C$ and $F$ anyons.

First we consider that the symmetry boundary is chosen by the pure charge condensation $A \oplus B \oplus 2C$, on which the topological lines generate the $S_{3}$ ($\text{Vec}_{S_{3}}$) symmetry. In this case, there are four physical boundaries corresponding to four possible gapped phases:
\begin{enumerate}
    \item $\mathcal{L}_{1} = A \oplus B \oplus 2C$ and $H=\{1\}$: $S_{3}$ SSB phase. In this case, the $S_3$ symmetry is completely broken, and there are 6 ground states.
    
    \item $\mathcal{L}_{2} = A \oplus D \oplus F$ and $H=S_3$: $S_{3}$ symmetric phase. There are no Wilson lines connecting the two boundaries, so the $S_3$ symmetry is unbroken. This is the $S_{3}$ symmetric state.
   
    \item $\mathcal{L}_{3} = A \oplus B \oplus 2F$ and $H=\z_3$: $Z_{2}$ SSB phase. The only nontrivial local operator is the local $B$ Wilson line, which braids trivially with the $r$ and $r^2$ lines on the symmetry boundary. Thus the remaining symmetry is $\z_3$. There are 2 ground states corresponding to the trivial local operator ${\cal O}_A$ and the nontrivial local operator ${\cal O}_B$, as expected from the $S_3\rightarrow \z_3$ SSB.
    
    \item $\mathcal{L}_{4} = A \oplus C \oplus D$ and $H=\z_2$: $Z_{3}$ SSB phase. The nontrivial local operators are given by ${\cal O}_C^{1}, {\cal O}_C^{2}$, where $1$ and $2$ label the condensation channel on the reference boundary (the channel is unique for $C$ on the right). They form an irreducible representation under the $S_3$ symmetry, so the remaining global symmetry is trivial. There are three ground states corresponding to the trivial local operators ${\cal O}_A$ and the other two nontrivial local operators ${\cal O}_C^{1}, {\cal O}_C^{2}$. The three ground states form a reducible representation, decomposing into the identity rep and the 2-dimensional rep.
\end{enumerate}

If we choose the symmetry boundary to be the flux condensed boundary $\mathcal{L}_{2} = A \oplus D \oplus F$, the topological lines generate the $\text{Rep}(S_{3})$ symmetry, which come from the charges $A,B,C$ in the bulk. The fusion rule of the $\text{Rep}(S_{3})$ symmetry is given by
\begin{align}
     X_{B} \otimes X_{B} &= X_{A}, \nonumber
     \\
     X_{B} \otimes X_{C} &= X_{C} \otimes X_{B} = X_{C}, \nonumber
     \\
     X_{C} \otimes X_{C} &= X_{A} \oplus X_{B} \oplus X_{C}.
\end{align}
The mapping from the boundary lines to the bulk anyon lines is given by
\begin{align}
    X_{A} &\rightarrow A \oplus D \oplus F, \nonumber
    \\
    X_{B} &\rightarrow B \oplus E \oplus F, \nonumber
    \\
    X_{C} &\rightarrow C \oplus D \oplus E \oplus G \oplus H.
\label{eq:Reps3bdybulkmap}
\end{align}

There are also four possible gapped phases by choosing the physical boundary to be one of the four gapped boundary of the $S_{3}$ gauge theory, which is studied in Ref.~\cite{Bhardwaj2023noninvertiblegapped1,Bhardwaj2023noninvertiblegapped2}. For the sake of completeness, we summarize the result below:
\begin{enumerate}
    \item $\mathcal{L}_{1} = A \oplus B \oplus 2C$: $\text{Rep}(S_{3})$ symmetric phase. In this case, the $\text{Rep}(S_{3})$ symmetry is preserved and there is a unique ground state.
    
    \item $\mathcal{L}_{2} = A \oplus D \oplus F$: $\text{Rep}(S_{3})$ SSB phase. There are three ground states, corresponding to the local operators: ${\cal O}_A, {\cal O}_D, {\cal O}_F$. Note that the local operators are formed by the pure fluxes, which are labeled by conjugacy classes $[g]$. The symmetry actions of the $\text{Rep}(S_{3})$ symmetry on these local operators is given by the characters:
    \begin{equation}
        U_{\rho} \circ {\cal O}_{[g]} = \chi_{\rho}(g) {\cal O}_{[g]}.
    \end{equation}
   
    \item $\mathcal{L}_{3} = A \oplus B \oplus 2F$: $\text{Rep}(S_{3})/\zz$ SSB phase. There are three ground states, corresponding to the local operators: ${\cal O}_A, {\cal O}_F^{1}, {\cal O}_F^{2}$, where ${\cal O}_F^{1}$ and ${\cal O}_F^{2}$ come from the two possible condensation channel on the physical boundary. The symmetry actions on the local operators are given by
    \begin{align}
        U_{B} &\circ {\cal O}_A = {\cal O}_A, \quad U_{B} \circ {\cal O}_F^{i} =  {\cal O}_F^{i}, \nonumber
        \\
        U_{C} &\circ {\cal O}_A = 2 {\cal O}_A, \quad U_{C} \circ {\cal O}_F^{i} =  -{\cal O}_F^{i},
    \end{align}
    where $i = 1,2$. 
    This implies that non-invertible symmetry $X_{C}$ is broken spontaneously but the $\zz$ symmetry is unbroken. 
    
    \item $\mathcal{L}_{4} = A \oplus C \oplus D$: $\zz$ SSB phase. The local operators are given by ${\cal O}_A, {\cal O}_D$ so that there are two ground states. The local operator ${\cal O}_D$ transformed non-trivially under both $X_{B}$ and $X_{C}$:
    \begin{equation}
        U_{B} \circ {\cal O}_D = - {\cal O}_D, \quad  U_{C} \circ {\cal O}_D = 0,
    \end{equation}
    which implies that the $\zz$ symmetry is broken spontaneously. 
    
\end{enumerate}

\subsection{Fermionization}
To perform the fermionization, we need to understand the fermionic gapped boundaries for the $S_{3}$ gauge theory. There are two fermionic gapped boundaries~\cite{Lou2021}:
\begin{align}
     \mathcal{L}_{1}^{f} &= A \oplus E \oplus F,
     \\
     \mathcal{L}_{2}^{f} &= A \oplus C \oplus E,
\end{align}
where the two fermionic boundaries are related by the automorphism $C \leftrightarrow F$.

We would like to fermionize the $\text{Rep}(S_{3})$ symmetry with respecting to the $\zz$ symmetry. To do so, we need the following $\zz$ twisted partition functions for systems with the $\text{Rep}(S_{3})$ symmetry:
\begin{align}
    Z_{0,0} = Z_{A} + Z_{D} + Z_{F}, \nonumber
    \\
    Z_{0,1} = Z_{A} - Z_{D} + Z_{F}, \nonumber
    \\
    Z_{1,0} = Z_{B} + Z_{E} + Z_{F}, \nonumber
    \\
    Z_{1,1} = Z_{B} - Z_{E} + Z_{F}. 
\end{align}
Applying the fermoinization formula Eq.~\eqref{eq:fermionization-2}, we find
\begin{equation}
    Z_{\mathcal{F}}[\text{A},\text{A}] = Z_{A} + Z_{E} + Z_{F},
\end{equation}
which means that the fermionization procedure is implemented by changing the symmetry boundary from $\mathcal{L}_{2} = A \oplus D \oplus F$ to $\mathcal{L}_{1}^{f} = A \oplus E \oplus F$.

The symmetry is generated by the line $X_{A}, X_{B_{f}}, X_{C}$, which satisfy the same fusion rule as the $\text{Rep}(S_{3})$ symmetry~\cite{Lou2021}:
\begin{align}
     X_{B_{f}} \otimes X_{B_{f}} &= X_{A}, \nonumber
     \\
     X_{B_{f}} \otimes X_{C} &= X_{C} \otimes X_{B_{f}} = X_{C}, \nonumber
     \\
     X_{C} \otimes X_{C} &= X_{A} \oplus X_{B_{f}} \oplus X_{C}.
\end{align}
Following Ref.~\cite{Inamura2023}, we refer to this super fusion category symmetry as $\text{SRep}(\mathcal{H}_{S_{3}})$\footnote{This super fusion category is equivalent to $\text{Rep}(S_{3}) \boxtimes \text{SVec}$. Ref.~\cite{Inamura2023} shows that in general fermionizing a $\text{Rep}(G)$ symmetry results in a super fusion category that is equivalent to $\text{Rep}(G) \boxtimes \text{SVec}$.}.

The fermion parity $\zz^{F}$ symmetry is generated by $X_{B_{f}}$. The mapping from the boundary lines to the bulk anyon lines is given by
\begin{align}
    X_{A} &\rightarrow A \oplus E \oplus F, \nonumber
    \\
    X_{B_{f}} &\rightarrow B \oplus D \oplus F, \nonumber
    \\
    X_{C} &\rightarrow C \oplus D \oplus E \oplus G \oplus H.
\label{eq:Reps3bdybulkmap2}
\end{align}
We will also choose their bulk representatives of $X_{A}$, $X_{B_{f}}$, and $X_{C}$ to be $A$, $B$, and $C$ respectively. Lattice models with $\text{SRep}(\mathcal{H}_{S_{3}})$ symmetry has been studied recently in Ref.~\cite{Chatterjee2024s3}.

\subsection{Gapped phases}
Here we discuss the possible gapped phases in fermionic chians with the super fusion category $\text{SRep}(\mathcal{H}_{S_{3}})$ symmetry, which are realized by choosing the physical boundary to be $\mathcal{L}_{i}$ for $i=1,2,3,4$.

\begin{enumerate}
    \item $\mathcal{L}_{1} = A \oplus B \oplus 2C$: $\text{SRep}(\mathcal{H}_{S_{3}})$ symmetric phase. In this case, the $\text{SRep}(\mathcal{H}_{S_{3}})$ symmetry is preserved and there is a unique ground state.
    
    \item $\mathcal{L}_{2} = A \oplus D \oplus F$: $\text{SRep}(\mathcal{H}_{S_{3}})/\zz$ SSB + Kitaev phase. There are two ground states, corresponding to the local operators: ${\cal O}_A$ and ${\cal O}_F$. The symmetry actions on the local operators are given by
    \begin{align}
        U_{B} &\circ {\cal O}_A = {\cal O}_A, \quad U_{B} \circ {\cal O}_F =  {\cal O}_F, \nonumber
        \\
        U_{C} &\circ {\cal O}_A = 2 {\cal O}_A, \quad U_{C} \circ {\cal O}_F =  -{\cal O}_F.
    \end{align} 
    This implies that non-invertible symmetry $X_{C}$ is broken spontaneously but the $\zz^{F}$ symmetry is unbroken. In particular, the system is in the Kitaev phase. This can be seen by calculating the partition function in the $(\text{P},\text{P})$ spin structure: 
    \begin{equation}
        Z[\text{P},\text{P}] = -Z_{D}-Z_{F} = -2,
    \end{equation}
    where the factor ``2" represents the 2-fold ground state degeneracy from the non-invertible symmetry breaking and the minus sign signals the Kitaev phase.
   
    \item $\mathcal{L}_{3} = A \oplus B \oplus 2F$: $\text{SRep}(\mathcal{H}_{S_{3}})/\zz$ SSB phase. There are three ground states, corresponding to the local operators: ${\cal O}_A, {\cal O}_F^{1}, {\cal O}_F^{2}$, where ${\cal O}_F^{1}$ and ${\cal O}_F^{2}$ come from the two possible condensation channel on the physical boundary. The symmetry actions on the local operators are given by
    \begin{align}
        U_{B} &\circ {\cal O}_A = {\cal O}_A, \quad U_{B} \circ {\cal O}_F^{i} =  {\cal O}_F^{i}, \nonumber
        \\
        U_{C} &\circ {\cal O}_A = 2 {\cal O}_A, \quad U_{C} \circ {\cal O}_F^{i} =  -{\cal O}_F^{i},
    \end{align}
    where $i = 1,2$. 
    This implies that non-invertible symmetry $X_{C}$ is broken spontaneously but the $\zz^{F}$ symmetry is unbroken.
    
    \item $\mathcal{L}_{4} = A \oplus C \oplus D$: $\text{SRep}(\mathcal{H}_{S_{3}})$ symmetric Kitaev phase. The $\text{SRep}(S_{3})$ symmetry is preserved and there is a unique ground state. However, the system is in the Kitaev phase since the partition function in the $(\text{P},\text{P})$ spin structure is non-trivial:
    \begin{equation}
        Z[\text{P},\text{P}] = -Z_{D} = -1,
    \end{equation}
    and the partition functions in the other spin structures are all equal to one.
\end{enumerate}

\subsection{Quantum critical points}

\begin{figure}
	\centering
	\includegraphics[width=0.48\textwidth]{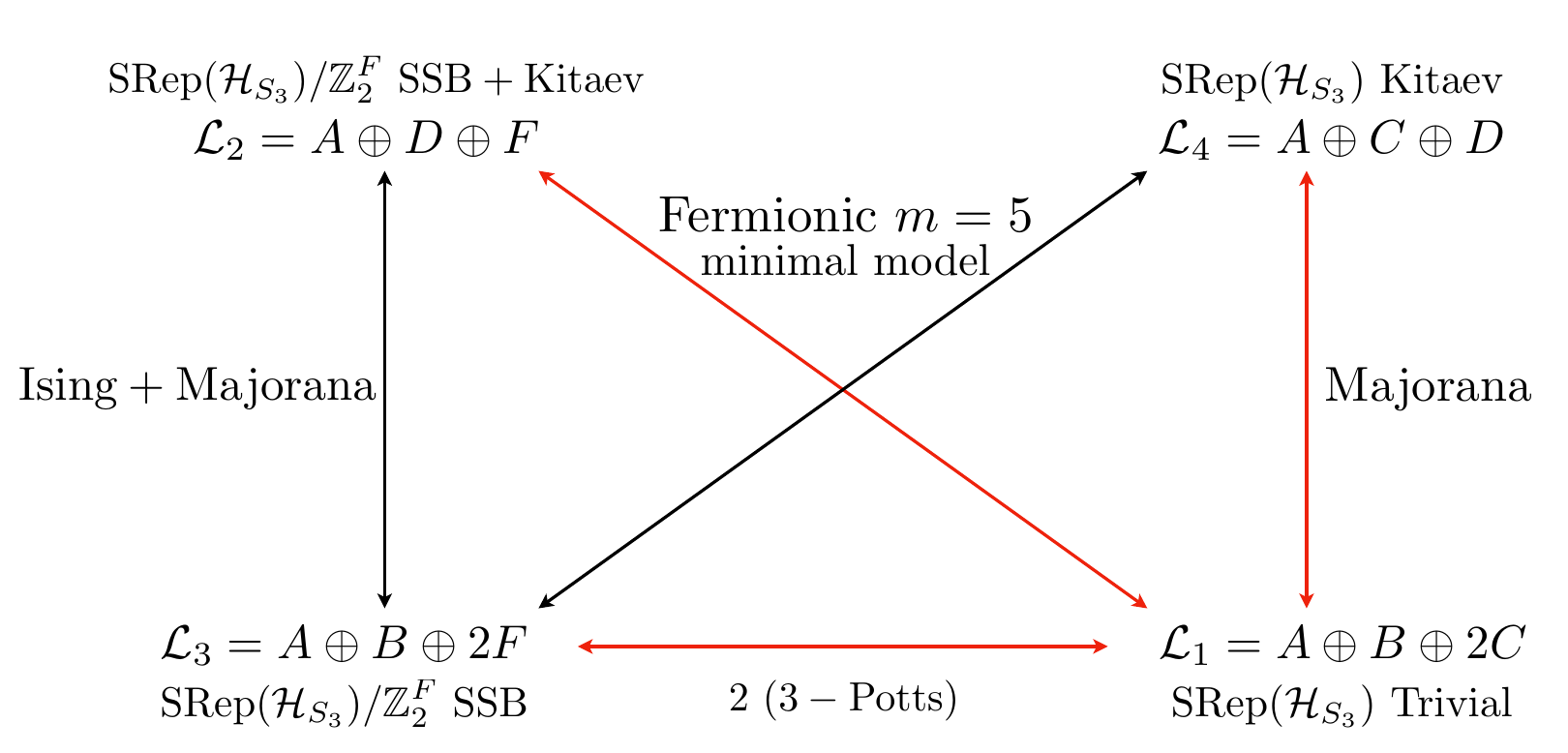}
	\caption{Summary of the phase transitions in the fermionic systems with $\text{SRep}(\mathcal{H}_{S_{3}})$ symmetry. The red arrows indicate the required input boundary critical theories on the physical boundary of the sandwich discussed in Sec.~\ref{sec:s3input}.}
	\label{fig:sreps3}
\end{figure}

Various quantum critical points between these fermionic gapped phases in the fermionic systems with $\text{SRep}(\mathcal{H}_{S_{3}})$ symmetry can be obtained by topological holography. We will begin with a review of the three boundary critical theories in Sec~\ref{sec:s3input}, which serve as our inputs to study the fermionic quantum critical points. By choosing the symmetry boundary to be appropriate bosonic gapped boundaries, we review how the sandwich construction produces the CFT descriptions of the phase transitions for the bosonic systems with $S_{3}$ and $\text{Rep}(S_{3})$ symmetry. We will proceed to discuss the fermionic critical theories in Sec.~\ref{sec:sreps31} - \ref{sec:sreps32}. The final result is summarized in Fig.~\ref{fig:sreps3}.

\subsubsection{Input critical theories}
\label{sec:s3input}

\paragraph{$S_{3}$ symmetric $\leftrightarrow$ $S_{3}$ SSB}

The first input we need is the sandwich picture for the critical point between the $S_{3}$ symmetric and $S_{3}$ SSB phases. In the sandwich picture, we tune the physical boundary to be sitting at the critical point between the gapped boundary $\mathcal{L}_{1}$ and $\mathcal{L}_{2}$ and the symmetry boundary is the charge condensed boundary $\mathcal{L}_{1}$. We denote the boundary critical theory as $\mathcal{P}_{12}$. It has been shown in Ref.~\cite{Chatterjee2023holo} that one of the possible CFT description for the boundary critical theory $\mathcal{P}_{12}$ is a CFT with central charge $c=4/5$. It has following multi-component partition functions in the anyon basis of the $S_{3}$ gauge theory written in terms of the characters of the $(6,5)$ minimal model:
\begin{align}
    Z_{A}^{(12)} &= |\chi_{0}|^{2} + |\chi_{3}|^{2} + |\chi_{\frac{2}{5}}|^{2} + |\chi_{\frac{7}{5}}|^{2}, \nonumber
    \\
    Z_{B}^{(12)} &= \chi_{0} \bar{\chi}_{3} + \chi_{3} \bar{\chi}_{0} + \chi_{\frac{2}{5}} \bar{\chi}_{\frac{7}{5}} + \chi_{\frac{7}{5}} \bar{\chi}_{\frac{2}{5}}, \nonumber
    \\
    Z_{C}^{(12)} &= |\chi_{\frac{2}{3}}|^{2} + |\chi_{\frac{1}{15}}|^{2}, \nonumber
    \\
    Z_{D}^{(12)} &= |\chi_{\frac{1}{8}}|^{2} + |\chi_{\frac{13}{8}}|^{2} + |\chi_{\frac{1}{40}}|^{2} + |\chi_{\frac{21}{40}}|^{2}, \nonumber
    \\
    Z_{E}^{(12)} &= \chi_{\frac{1}{8}} \bar{\chi}_{\frac{13}{8}} + \chi_{\frac{13}{8}} \bar{\chi}_{\frac{1}{8}} + \chi_{\frac{1}{40}} \bar{\chi}_{\frac{21}{40}} + \chi_{\frac{21}{40}} \bar{\chi}_{\frac{1}{40}}, \nonumber
    \\
    Z_{F}^{(12)} &= |\chi_{\frac{2}{3}}|^{2} + |\chi_{\frac{1}{15}}|^{2}, \nonumber 
    \\
    Z_{G}^{(12)} &= \chi_{0} \bar{\chi}_{\frac{2}{3}} + \chi_{3} \bar{\chi}_{\frac{2}{3}} + \chi_{\frac{2}{5}} \bar{\chi}_{\frac{1}{15}} + \chi_{\frac{7}{5}} \bar{\chi}_{\frac{1}{15}}, \nonumber
    \\
    Z_{H}^{(12)} &= \chi_{\frac{2}{3}} \bar{\chi}_{0} + \chi_{\frac{2}{3}} \bar{\chi}_{3} + \chi_{\frac{1}{15}} \bar{\chi}_{\frac{2}{5}} + \chi_{\frac{1}{15}} \bar{\chi}_{\frac{7}{5}}.
\label{eqn:Z65anyon1}
\end{align}

Since the symmetry boundary is given by $\mathcal{L}_{1} = A \oplus B \oplus 2C$, we have the partition function 
\begin{align}
     Z &= Z_{A}^{(12)} + Z_{B}^{(12)} + 2 Z_{C}^{(12)} \nonumber
     \\
     &= |\chi_{0} + \chi_{3}|^{2} + |\chi_{\frac{2}{5}} + \chi_{\frac{7}{5}}|^{2} + 2 |\chi_{\frac{2}{3}}|^{2} + 2|\chi_{\frac{1}{15}}|^{2},
\end{align}
which is the of the partition function of the 3-state Potts model. 

We note that if we change the symmetry boundary to be $\mathcal{L}_{2} = A \oplus D \oplus F$, the physical boundary $\mathcal{L}_{1}$ then corresponds to the $\text{Rep}(S_{3})$ symmetric phase and $\mathcal{L}_{2}$ corresponds to the $\text{Rep}(S_{3})$ SSB phase. Now we tune the physical boundary to be described by the theory $\mathcal{P}_{12}$ Eq.~\eqref{eqn:Z65anyon1}, the transition between the $\text{Rep}(S_{3})$ symmetric and the $\text{Rep}(S_{3})$ SSB phases is given by
\begin{widetext}
\begin{align}
     Z &= Z_{A}^{(12)} + Z_{D}^{(12)} + Z_{F}^{(12)} \nonumber
     \\
     &= |\chi_{0}|^{2} + |\chi_{3}|^{2} + |\chi_{\frac{2}{5}}|^{2} + |\chi_{\frac{7}{5}}|^{2} + |\chi_{\frac{1}{8}}|^{2} + |\chi_{\frac{13}{8}}|^{2} + |\chi_{\frac{1}{40}}|^{2} + |\chi_{\frac{21}{40}}|^{2} + |\chi_{\frac{2}{3}}|^{2} + |\chi_{\frac{1}{15}}|^{2},
\end{align}
which is the partition function of the tetracritical Ising model. 
\end{widetext}

\paragraph{$Z_{2}$ SSB $\leftrightarrow$ $S_{3}$ SSB}
The second input is the sandwich picture for the transition between the $Z_{2}$ SSB and $S_{3}$ SSB phases. The physical boundary is at the transition between $\mathcal{L}_{1}$ and $\mathcal{L}_{3}$. We denote the boundary critical theory as $\mathcal{P}_{13}$. Ref.~\cite{Chatterjee2023holo} found that it can be described by the following multi-component partition functions:
\begin{align}
    Z_{A}^{(13)} &= |\chi_{0} + \chi_{3}|^{2} + |\chi_{\frac{2}{5}} + \chi_{\frac{7}{5}}|^{2}, \nonumber
    \\
    Z_{B}^{(13)} &= |\chi_{0} + \chi_{3}|^{2} + |\chi_{\frac{2}{5}} + \chi_{\frac{7}{5}}|^{2}, \nonumber
    \\
    Z_{C}^{(13)} &= 2 (|\chi_{\frac{2}{3}}|^{2} +|\chi_{\frac{1}{15}}|^{2}), \nonumber
    \\
    Z_{D}^{(13)} &= 0, \nonumber
    \\
    Z_{E}^{(13)} &= 0, \nonumber
    \\
    Z_{F}^{(13)} &= 2 ( |\chi_{\frac{2}{3}}|^{2} +|\chi_{\frac{1}{15}}|^{2}), \nonumber 
    \\
    Z_{G}^{(13)} &= \chi_{0} \bar{\chi}_{\frac{2}{3}} + \chi_{3} \bar{\chi}_{\frac{2}{3}} + \chi_{\frac{2}{5}} \bar{\chi}_{\frac{1}{15}} + \chi_{\frac{7}{5}} \bar{\chi}_{\frac{1}{15}}, \nonumber
    \\
    Z_{H}^{(13)} &= \chi_{\frac{2}{3}} \bar{\chi}_{0} + \chi_{\frac{2}{3}} \bar{\chi}_{3} + \chi_{\frac{1}{15}} \bar{\chi}_{\frac{2}{5}} + \chi_{\frac{1}{15}} \bar{\chi}_{\frac{7}{5}}.
\label{eqn:Z65anyon2}
\end{align}
The partition function for the $Z_{2}$ SSB $\leftrightarrow$ $S_{3}$ SSB transition is then given by
\begin{align}
     Z &= Z_{A}^{(13)} + Z_{B}^{(13)} + 2 Z_{C}^{(13)} \nonumber
     \\
     &= 2 ( |\chi_{0} + \chi_{3}|^{2} + |\chi_{\frac{2}{5}} + \chi_{\frac{7}{5}}|^{2} + 2 |\chi_{\frac{2}{3}}|^{2} +2|\chi_{\frac{1}{15}}|^{2}),
\end{align}
where the prefactor ``2" in front of the parenthesis corresponds the 2-fold ground state degeneracy since the $\zz$ symmetry has been broken spontaneously. The transition is given by 2 copies of the 3-state Potts model. 

We note that we can also obtained the transition between the $\text{Rep}(S_{3})$ symmetric and $\text{Rep}(S_{3})/\zz$ SSB phases, at which the partition function is given by summing over the pure fluxes $Z = Z_{A}^{(13)} + Z_{D}^{(13)} + Z_{F}^{(13)}$. One can check that it equals to the partition function of the 3-state Potts model. 

\paragraph{$Z_{3}$ SSB $\leftrightarrow$ $S_{3}$ SSB}
The third input is the transition between the $Z_{3}$ SSB and $S_{3}$ SSB phases. The physical boundary is tuned to be sitting at the transition between $\mathcal{L}_{1}$ and $\mathcal{L}_{4}$. We denote the boundary critical theory as $\mathcal{P}_{14}$. Ref.~\cite{Chatterjee2023holo} found that it can be described by the following multi-component partition functions in terms of the characters of the Ising CFT:
\begin{align}
    Z_{A}^{(14)} &= |\chi_{1}|^{2} + |\chi_{\psi}|^{2}, \nonumber
    \\
    Z_{B}^{(14)} &= |\chi_{\sigma}|^{2},  \nonumber
    \\
    Z_{C}^{(14)} &= |\chi_{1}|^{2} + |\chi_{\sigma}|^{2} + |\chi_{\psi}|^{2},   \nonumber
    \\
    Z_{D}^{(14)} &= |\chi_{\sigma}|^{2},  \nonumber
    \\
    Z_{E}^{(14)} &= \chi_{1}\bar{\chi}_{\psi} + \chi_{\psi}\bar{\chi}_{1}, \nonumber
    \\
    Z_{F}^{(14)} &= 0, \nonumber 
    \\
    Z_{G}^{(14)} &= 0, \nonumber
    \\
    Z_{H}^{(14)} &= 0.
\label{eqn:Z65anyon3}
\end{align}
The partition function for the $Z_{3}$ SSB $\leftrightarrow$ $S_{3}$ SSB transition is given by
\begin{align}
     Z &= Z_{A}^{(14)} + Z_{B}^{(14)} + 2 Z_{C}^{(14)} \nonumber
     \\
     &= 3 (|\chi_{1}|^{2} + |\chi_{\sigma}|^{2}+ |\chi_{\psi}|^{2}),
\end{align}
which is three copies of the Ising CFT as expected since the $\z_{3}$ symmetry has been broken spontaneously. 

We can also obtained the transition between the $\text{Rep}(S_{3})$ symmetric and $\zz$ SSB phases. The partition function is given by summing over the pure fluxes $Z = Z_{A}^{(14)} + Z_{D}^{(14)} + Z_{F}^{(14)}$, which is just a single copy of the Ising CFT. 

\subsubsection{$\text{SRep}(\mathcal{H}_{S_{3}})$ Trivial $\leftrightarrow$ ($\text{SRep}(\mathcal{H}_{S_{3}})/\zz$ SSB + Kitaev) = fermionic minimal model $(m=5)$}
\label{sec:sreps31}

Now we begin the discussion of the phase transitions in fermionic systems with $\text{SRep}(\mathcal{H}_{S_{3}})$ symmetry. Recall that the $\text{SRep}(\mathcal{H}_{S_{3}})$ symmetric trivial phase corresponds to the physical boundary $\mathcal{L}_{1}$ and the ($\text{SRep}(\mathcal{H}_{S_{3}})/\zz$ + Kitaev) phase corresponds to the physical boundary $\mathcal{L}_{2}$. We tune the physical boundary to be described by the critical theory $\mathcal{P}_{12}$ Eq.~\eqref{eqn:Z65anyon1}. We perform the fermionization by changing the symmetry boundary to $\mathcal{L}_{1}^{f} = A \oplus E \oplus F$. The partition function in the $(\text{A},\text{A})$ spin structure is 
\begin{widetext}
\begin{align}
     Z[\text{A},\text{A}] &= Z_{A}^{(12)} + Z_{E}^{(12)} + Z_{F}^{(12)} \nonumber
     \\
     &= |\chi_{0}|^{2} + |\chi_{3}|^{2} + |\chi_{\frac{2}{5}}|^{2} + |\chi_{\frac{7}{5}}|^{2} +
     + |\chi_{\frac{2}{3}}|^{2} + |\chi_{\frac{1}{15}}|^{2}+
     \chi_{\frac{1}{8}} \bar{\chi}_{\frac{13}{8}} + \chi_{\frac{13}{8}} \bar{\chi}_{\frac{1}{8}} + \chi_{\frac{1}{40}} \bar{\chi}_{\frac{21}{40}} + \chi_{\frac{21}{40}} \bar{\chi}_{\frac{1}{40}},
\label{eqn:Zfm5}
\end{align}
\end{widetext}
which is precisely the partition function of the fermionic $m=5$ minimal model in the $(\text{A},\text{A})$ spin structure~\cite{Hsieh2021}. 

\subsubsection{$\text{SRep}(\mathcal{H}_{S_{3}})$ Kitaev $\leftrightarrow$  $\text{SRep}(\mathcal{H}_{S_{3}})/\zz$ SSB = fermionic minimal model $(m=5)$}

We now discuss the $\text{SRep}(\mathcal{H}_{S_{3}})$ Kitaev $\leftrightarrow$  $\text{SRep}(\mathcal{H}_{S_{3}})/\zz$ SSB transition. The $\text{SRep}(\mathcal{H}_{S_{3}})$ Kitaev phase corresponds to the physical boundary $\mathcal{L}_{4}$ and the $\text{SRep}(\mathcal{H}_{S_{3}})/\zz$ SSB phase corresponds to the physical boundary $\mathcal{L}_{2}$. We denote the boundary critical theory as $\mathcal{P}_{24}$ and the corresponding critical boundary on the sandwich as $\ket{\psi_{24}}$. Due to the automorphism in the $S_{3}$ gauge theory that exchanges $C$ and $F$ anyons, the boundary critical theory as $\mathcal{P}_{24}$ is related to the $\mathcal{P}_{13}$ by an insertion of the $C$-$F$-exchanging defect $\mathcal{D}_{a}$. The partition function is thus given by the overlap:
\begin{align}
    Z[\text{A},\text{A}] &= \ip{\mathcal{L}_{1}^{f}}{\psi_{24}} \nonumber
    \\
    &= \me{\mathcal{L}_{1}^{f}}{\mathcal{D}_{a}}{\psi_{13}} \nonumber
    \\
    &= \ip{\mathcal{L}_{2}^{f}}{\psi_{13}} \nonumber
    \\
    &= Z_{A}^{(13)} + Z_{C}^{(13)} + Z_{E}^{(13)},
\end{align}
which produces the same partition function of the fermionic $m=5$ minimal model in Eq.~\eqref{eqn:Zfm5}. Therefore, we found that the following two transitions:
\begin{align}
    \text{SRep}(\mathcal{H}_{S_{3}}) \ \text{Trivial} &\leftrightarrow \text{SRep}(\mathcal{H}_{S_{3}})/\zz \ \text{SSB} + \text{Kitaev}, \nonumber
    \\
    \text{SRep}(\mathcal{H}_{S_{3}}) \ \text{Kitaev} &\leftrightarrow \text{SRep}(\mathcal{H}_{S_{3}})/\zz \ \text{SSB},
\end{align}
are both described by the same fermionic $m=5$ minimal model, which is the direct consequence of the $C$-$F$-exchanging automorphism. 

\subsubsection{$\text{SRep}(\mathcal{H}_{S_{3}})$ Trivial $\leftrightarrow$  $\text{SRep}(\mathcal{H}_{S_{3}})/\zz$ SSB = 3-Potts $\times$ $\mathcal{F}_{0}$}

Here we discuss the transition between the $\text{SRep}(\mathcal{H}_{S_{3}})$ symmetric trivial and the $\text{SRep}(\mathcal{H}_{S_{3}})/\zz$ SSB phases. Since correspond to the gapped boundaries $\mathcal{L}_{1}$ and $\mathcal{L}_{3}$ respectively, the transition between them can be described by the critical theory $\mathcal{P}_{13}$ Eq.~\eqref{eqn:Z65anyon2}. The partition function in the $(\text{A},\text{A})$ spin structure is 
\begin{align}
     Z &= Z_{A}^{(13)} + Z_{E}^{(13)} +  Z_{F}^{(13)} \nonumber
     \\
     &= |\chi_{0} + \chi_{3}|^{2} + |\chi_{\frac{2}{5}} + \chi_{\frac{7}{5}}|^{2} + 2 |\chi_{\frac{2}{3}}|^{2} + 2|\chi_{\frac{1}{15}}|^{2},
\end{align}
which is the 3-state Potts model stacking with a trivial fermion theory $\mathcal{F}_{0}$.

\subsubsection{$\text{SRep}(\mathcal{H}_{S_{3}})$ Trivial $\leftrightarrow$  $\text{SRep}(\mathcal{H}_{S_{3}})$ Kitaev = Majorana}

We proceed to discuss the transition between the $\text{SRep}(\mathcal{H}_{S_{3}})$ symmetric trivial and  $\text{SRep}(\mathcal{H}_{S_{3}})$ symmetric Kitaev phases. They correspond to the gapped boundaries $\mathcal{L}_{1}$ and $\mathcal{L}_{4}$ respectively. The transition between them can be described by the critical theory $\mathcal{P}_{14}$ Eq.~\eqref{eqn:Z65anyon3}. The partition function in the $(\text{A},\text{A})$ spin structure is 
\begin{align}
     Z &= Z_{A}^{(14)} + Z_{E}^{(14)} +  Z_{F}^{(14)} \nonumber
     \\
     &= |\chi_{1}|^{2} + |\chi_{\psi}|^{2} + \chi_{1}\bar{\chi}_{\psi} + \chi_{\psi}\bar{\chi}_{1}.
\end{align}
We recognize it is the free Majorana fermion CFT.

\subsubsection{$\text{SRep}(\mathcal{H}_{S_{3}})/\zz$ SSB $\leftrightarrow$ ($\text{SRep}(\mathcal{H}_{S_{3}})/\zz$ SSB + Kitaev) = Ising + Majorana}
\label{sec:sreps32}

Finally we discuss the transition between the $\text{SRep}(\mathcal{H}_{S_{3}})/\zz$ SSB and the ``$\text{SRep}(\mathcal{H}_{S_{3}})/\zz$ SSB + Kitaev" phases. They correspond to the gapped boundaries $\mathcal{L}_{2}$ and $\mathcal{L}_{3}$ respectively. Denoting the boundary critical theory as $\mathcal{P}_{23}$, it's related to the critical theory $\mathcal{P}_{14}$ by the $C$-$F$-exchanging defect $\mathcal{D}_{a}$. The partition function is then given by the overlap:
\begin{widetext}
\begin{align}
    Z[\text{A},\text{A}] &= \ip{\mathcal{L}_{1}^{f}}{\psi_{23}} \nonumber
    \\
    &= \me{\mathcal{L}_{1}^{f}}{\mathcal{D}_{a}}{\psi_{14}} \nonumber
    \\
    &= \ip{\mathcal{L}_{2}^{f}}{\psi_{14}} \nonumber
    \\
    &= Z_{A}^{(14)} + Z_{C}^{(14)} + Z_{E}^{(14)}
    \\
    &= ( |\chi_{1}|^{2} + |\chi_{\sigma}|^{2} + |\chi_{\psi}|^{2} ) + ( |\chi_{1}|^{2} + |\chi_{\psi}|^{2} + \chi_{1}\bar{\chi}_{\psi} + \chi_{\psi}\bar{\chi}_{1} ),
\end{align}
\end{widetext}
which is the ``sum" of the Ising and the Majorana CFTs. There are two superselection sectors: one sector comprises an Ising CFT and the other sector a Majorana CFT. This is an example of a \emph{gapless SSB} phase~\cite{Bhardwaj2024hasse}. The gapless SSB phase in this example is fermionic in the sense that the spontaneously broken non-invertible $X_{C}$ symmetry causing two superselection sectors where one sector is bosonic and the other sector is fermionic. This concludes the discussion of the phase transitions in the fermioic systems with $\text{SRep}(\mathcal{H}_{S_{3}})$ symmetry. The final result is summarized in Fig.~\ref{fig:sreps3}.

\section{Discussion}
\label{sec:discussion}

In this work, we apply the concept of topological holography to study fermionic gapped phases and phase transitions. Potential CFT descriptions for the phase transitions are proposed based on this framework. We consider a generalization of the sandwich construction tailored to fermionic systems such that the symmetry boundary is a fermionic gapped boundary condition of the SymTFT. In particular, performing the fermionization corresponds to changing the symmetry boundary from a bosonic gapped boundary to a fermionic one. We establish connections between the formal fermionization formula and the choice of fermionic gapped boundary conditions. We also have discuss the encoding of spin structures within the SymTFT framework. 

We demonstrate the main idea through various examples, including the fermionic systems with $\zz^{F}$, $\zz \times \zz^{F}$, $\z_{4}^{F}$, and the non-invertible super fusion category $\text{SRep}(\mathcal{H}_{S_{3}})$ symmetry. We find many exotic fermionic quantum critical points and gapless phases, including two kinds of fermionic SEQCPs in the $\zz \times \zz^{F}$ fermionic chains. The first kind of the fermionic SEQCPs is obtained by stacking with a fermionic SPT states ($\textbf{T}$ operation), and the other new type of the fermionic SEQCPs is obtained by a twisted gauging ($\textbf{T}\textbf{S}$ operation). We also find a fermionic gapless SPT phase with $\z_{4}^{F}$ symmetry, and a fermionic gapless SSB phase that breaks the fermionic non-invertible symmetry in $\text{SRep}(\mathcal{H}_{S_{3}})$ symmetry. 

This work opens up many interesting avenues for future exploration. Firstly, it would be interesting to study the fermionic SEQCPs obtained by the twisted gauging more generally. It would also be interesting to apply the topological holography to systematically study the fermionic quantum critical points and gapless phases with emergent anomalies. We expect the general classification of the \spd{2} invertible fermionic topological phases with symmetry~\cite{Aasen2021,Barkeshli2022finv,Ning2023} would be helpful in this direction. 

In this work, we only consider a simple example of the fermionic non-invertible symmetry obtained by fermionizing the $\zz$ subgroup of the $\text{Rep}(S_{3})$ symmetry. A general framework of the fermionization of fusion category symmetries in \spd{1} has been developed in Ref.~\cite{Inamura2023}. It would be interesting to study the fermionic phases with non-invertible symmetry and their phase transitions more systematically from the topological holography picture. 

Finally, the topological holography framework for fermionic systems has been formally developed in higher dimensions~\cite{Freed2022,Debray2023}. To apply this framework to physically relevant systems, it is essential to enhance our understanding of the physical picture and develop computational tools for fermionic gapped boundaries in higher-dimensional SymTFTs.

\begin{acknowledgements}
The author thanks Ryohei Kobayashi for the initial participation of this work and for many enlightening discussions, Meng Cheng for a previous collaboration and helpful discussions on topological holography, and Shu-Heng Shao for helpful discussions. 

\emph{Note added}: while this work was being finalized for posting, we become aware of an independent work that has some overlap with our results~\cite{Wen:2024udn}. We thank Lakshya Bhardwaj, Kansei Inamura, and Apoorv Tiwari for coordinating submission of their related independent work~\cite{Bhardwaj2024fsymtft}. 
\end{acknowledgements}

\appendix

\section{Fermionic gapped boundaries and super Lagrangian algebra}
\label{app:superL}
We give a brief review of the fermionic gapped boundaries and the super-Lagrangian algebra in thie appendix. We begin with a review of the Lagrangian algebra in a MTC $\mathcal{C}$. 

An algebra in a MTC $\mathcal{C}$ is a direction sum of of simple objects: $\mathcal{A} = \oplus_{\alpha} w_{\alpha} \alpha$, $\alpha \in \mathcal{C}$. More formally, we define a Lagrangian algebra $\mathcal{A}$ in a MTC $\mathcal{C}$ to be an object $\mathcal{A} \in \mathcal{C}$ along with a multiplication morphism $\mu : \mathcal{A} \otimes \mathcal{A} \rightarrow \mathcal{A}$ and unit morphism $\iota: 1 \rightarrow \mathcal{A}$ such that the following conditions hold:
\begin{itemize}
    \item Commutativity: $\mu \circ P_{\cal{A},\cal{A}} = \mu$, where $P_{\cal{A},\cal{A}}$ is the braiding in $\mathcal{C}$. It can be expressed diagrammatically:
\begin{equation}
    \includegraphics[width=.25\linewidth]{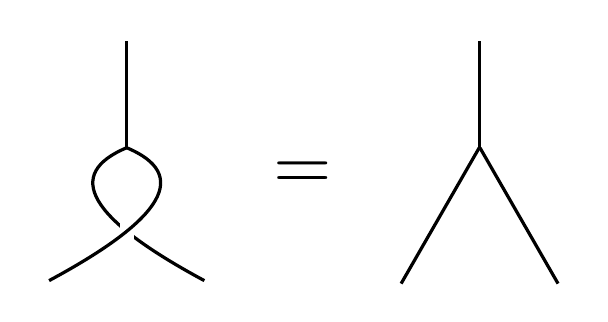},
\end{equation}
    where solid line represents $\mathcal{A}$ and the junction where three $\mathcal{A}$ lines meet is the morphism $\mu$.
    
    \item Associativity: $\mu \circ (\mu \otimes id) = \mu \circ (id \otimes \mu)$,
\begin{equation}
    \includegraphics[width=.3\linewidth]{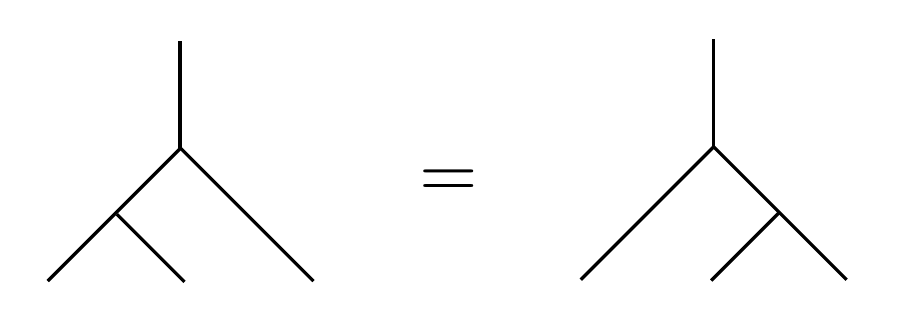}.
\end{equation}

    \item Unit: $\mu \circ (\iota \otimes id) = id$,
\begin{equation}
    \includegraphics[width=.15\linewidth]{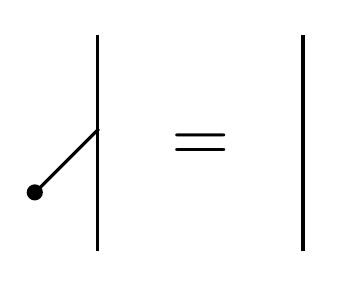}.
\end{equation}
$\mathcal{A}$ is called connected if $\text{Hom}(1,\mathcal{A})$ is 1-dimensional

    \item Separability: There exists a splitting morphism $\sigma : \mathcal{A} \rightarrow \mathcal{A}  \otimes \mathcal{A}$ such that $\mu \circ \sigma = id$, and satisfies
\begin{equation}
    \includegraphics[width=.5\linewidth]{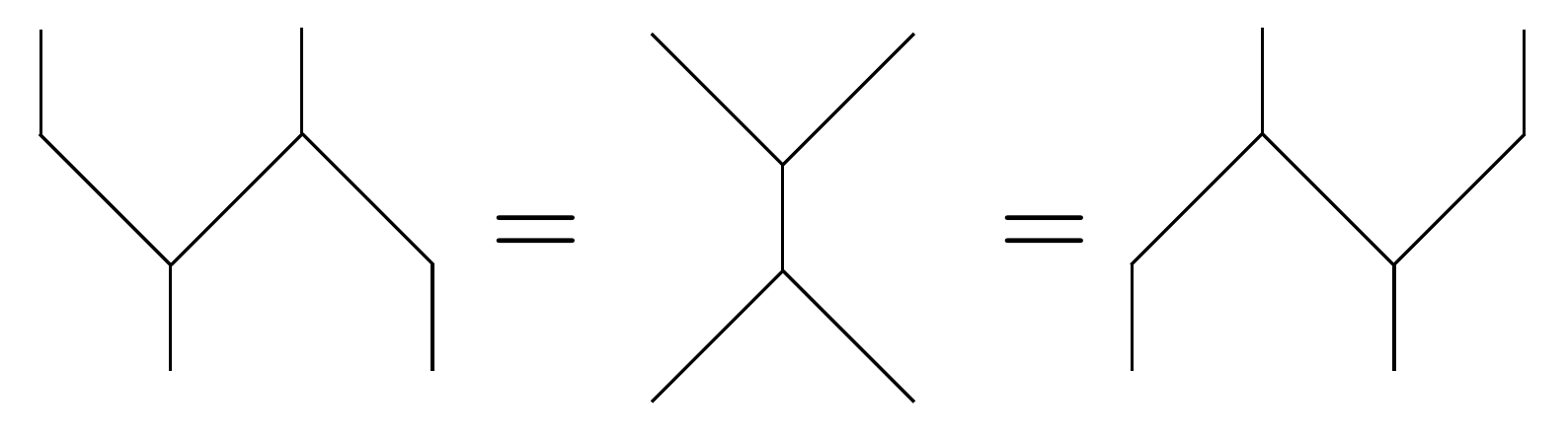}.
\end{equation}

    \item Lagrangian:
    $D_{\mathcal{A}} = D_{\mathcal{C}}$,
    where $D_{\mathcal{A}} = \sum_{\alpha \in \mathcal{C}} w_{\alpha} d_{\alpha}$ is the dimension of the algebra $\mathcal{A} = \oplus_{\alpha} w_{\alpha} \alpha$, and $D_{\mathcal{C}} = \sqrt{\sum_{\alpha \in \mathcal{C}} d_{\alpha}^{2}}$ is the total quantum dimension of $\mathcal{C}$.
\end{itemize}

We follow Ref.~\cite{Lou2021} to describe the fermionic gapped boundaries by relaxing the commutative condition to ``super-commutativity". A super algebra is a graded algebra by the $\zz^{F}$ symmetry:
\begin{equation}
    \mathcal{A}=\mathcal{A}_{0} \oplus \mathcal{A}_{1}
\end{equation}
Fermion parity $\sigma(i) = (-1)^{p}$ is assigned to each anyon labeled by $i$ in $\mathcal{A}_{p}$. The super-commutativity proposed in Ref.~\cite{Lou2021} is given by
\begin{equation}
    \mu \circ P_{c_{i},c_{j}} (-1)^{\sigma(c_{i}),\sigma(c_{ij)}} = \mu , \quad c_{i},c_{j} \in \mathcal{A},
\end{equation}
where $P_{c_{i},c_{j}}$ is the half-braid of the MTC $\mathcal{C}$. In this work, we call a Lagrangian algebra with the $\zz^{F}$ graded structure and the super-commutativity a super-Lagrangian algebra.

\section{Fermionic minimal models}
\label{app:fminimal}  

\begin{table*}[t]\centering
	\begin{tabular}{c|c|c}
		 & untwisted & twisted
		\\
		\hline
		even & S & U
		\\ 
		\hline
		odd & T & V
	\end{tabular}
	\caption{States of a bosonic theory on a circle with untwisted or twisted boundary condition can be arranged into four sectors.}
	\label{Table:bosonicsectors}
\end{table*}

\begin{table*}[t]\centering
	\begin{tabular}{c|c|c}
		 & A & P
		\\
		\hline
		bosonic & S & T
		\\ 
		\hline
		fermionic & V & U
	\end{tabular}
	\caption{States of a fermionic theory on a circle with antiperiodic (A) or periodic (P) boundary condition can be arranged into four sectors.}
	\label{Table:fermionicsectors}
\end{table*}

In this appendix, we discuss the sandwich picture for the fermionic minimal models. Fermionic minimal models are obtained by fermionizing a non-anomalous $\zz$ symmetry in the bosonic minimal models. In the sandwich construction, we can choose the bulk to be a $\zz$ toric code topological order and the physical boundary to be a bosonic minimal model. If the symmetry boundary is chosen to be a bosonic gapped boundary, say the $e$-condensed boundary, the sandwich construction realizes the bosonic minimal models. We fermionize the system by changing the symmetry boundary to be a fermion $f$-condensed boundary.

A bosonic theory with a non-anomalous $\zz$ symmetry on a circle $S^{1}$ can be either untwisted or twisted by the $\zz$ symmetry around $S^{1}$. The states in the Hilbert space in the untwisted and the twisted systems can each be decomposed into even and odd sectors under the $\zz$ symmetry. This is summarized in Table.~\ref{Table:bosonicsectors}, where $S$, $T$, $U$ and $V$ labels the states in the corresponding sectors. We will abuse the notation and use them to also denote the corresponding sectors in the partition functions. From the sandwich picture, we have the relation between the twisted partition functions and the multi-component partition functions in the toric code anyon basis:
\begin{align}
    Z_{0,0} &= Z_{1} + Z_{e}, \nonumber
    \\
    Z_{0,1} &= Z_{1} - Z_{e}, \nonumber
    \\
    Z_{1,0} &= Z_{m} + Z_{f}, \nonumber
    \\
    Z_{1,1} &= Z_{m} - Z_{f}.
\label{eqn:z2ztwist}
\end{align}
Using Eq.~\eqref{eqn:z2ztwist}, we have the following correspondence
\begin{align}
    S &\leftrightarrow Z_{1}, \quad T \leftrightarrow Z_{e}, \nonumber
    \\
    U &\leftrightarrow Z_{m}, \quad V \leftrightarrow Z_{f}.
\label{eqn:seccor}
\end{align}

After we fermionize the bosonic theory, the fermionic theory can be periodic or antiperiodic. The states in the Hilbert space in  systems with periodic or antiperiodic boundary condition can each be decomposed into even and odd sectors under the $\zz^{F}$ symmetry. In particular, the effect of the fermoinization is to permute the twisted sectors, which is summarized in Table.~\ref{Table:fermionicsectors}. From the sandwich picture, we have the relations between the partition functions in different spin structures and the partition functions in the toric code anyon basis:
\begin{align}
    Z[\text{A},\text{A}] &= Z_{1} + Z_{f}, \nonumber
    \\
    Z[\text{A},\text{P}] &= Z_{1} - Z_{f}, \nonumber
    \\
    Z[\text{P},\text{A}] &= Z_{e} + Z_{m}, \nonumber
    \\
    Z[\text{P},\text{P}] &= Z_{e} - Z_{m}.
\label{eqn:z2zspin}
\end{align}
Using the correspondence Eq.~\eqref{eqn:seccor} and the explicitly expressions for each sector in terms of the characters of the minimal models\cite{Hsieh2021}, one can obtain the partition functions in different spin structures for the fermionic minimal models by using Eq.~\eqref{eqn:z2zspin}. For the sake of completeness, here we give the explicitly expressions for each sector~\cite{Hsieh2021}. 
\begin{align}
    Z_{1} &= \sum_{r \equiv 1} \sum_{s} \chi_{r,s} \overline{\chi}_{r,s},
    \\
    Z_{e} &= \sum_{r \equiv 0} \sum_{s} \chi_{r,s} \overline{\chi}_{r,s},
    \\
    Z_{m} &= \sum_{r \equiv \frac{q}{2}} \sum_{s} \chi_{r,s} \overline{\chi}_{q-r,s},
    \\
    Z_{f} &= \sum_{r \equiv \frac{q}{2}+1} \sum_{s} \chi_{r,s} \overline{\chi}_{q-r,s},
\end{align}
where we denote Virasoro characters by $\chi_{r,s}$ with central charge $c=1-\frac{6}{m(m+1)}$, and $a \equiv b$ means the equality modulo 2. The range of $r$ and $s$ are given by $1 \leq r \leq q-1$ and $1 \leq s \leq p-1$, where $q=m$ and $p=m+1$ for even $m$, and $q=m+1$ and $p=m$ for odd $m$. Note that there is a redundancy for the characters $\chi_{r,s} = \chi_{q-r,p-s}$. This redundancy can be removed by the restriction $s \leq (p-1)/2$.


\bibliography{fsymtft,BCFT}


\end{document}